\documentclass[journal,twocolumn]{IEEEtran}

\IEEEoverridecommandlockouts

\usepackage{graphics} 
\usepackage{epsfig} 
\usepackage{times} 
\usepackage{amsmath} 
\usepackage{amssymb}  
\usepackage{multirow}
\usepackage[ruled,linesnumbered]{algorithm2e}
\usepackage{epstopdf} 
\usepackage{subfigure} 
\usepackage{stfloats} 
\usepackage{bm}
\usepackage{color}
\usepackage{amsthm}
\usepackage{cite}
\usepackage{array}
\usepackage{float}
\usepackage{mathrsfs}
\newcommand{\todo}[1]{\textcolor{red}{\textbf{TODO: {#1}}}}

\usepackage{nomencl}
\usepackage[switch]{lineno}

\makenomenclature

\begin{document}

\title{Machine Learning for Estimation and Control of Quantum Systems}
	
\author{Hailan Ma, Bo Qi, Ian R. Petersen, Re-Bing Wu, Herschel Rabitz, Daoyi Dong

 \thanks{This work was supported by the Australian Research Council’s Future Fellowship Funding Scheme under Project FT220100656 and the Discovery Project Funding Scheme under Project DP240101494.}%
\thanks{H. Ma, I. R. Petersen, and D. Dong are with the School of  Engineering, Australian National University, Canberra, ACT 2601, Australia (email: hailanma0413@gmail.com, i.r.petersen@gmail.com, daoyi.dong@anu.edu.au).}
	\thanks{B. Qi is with the Key Laboratory of Systems and Control, Academy of Mathematics and Systems Science, Chinese Academy of Sciences, Beijing, 100190, China (email: qibo@amss.ac.cn).}
 \thanks{R.-B. Wu is with the Center for Intelligent and Networked Systems,  Department of Automation, Tsinghua University, Beijing, 100084, China (email: rbwu@tsinghua.edu.cn).}
	\thanks{H.  Rabitz is with the  Department of  Chemistry,  Princeton University,    Princeton, NJ  08544, USA (email: hrabitz@princeton.edu).}
   }

	\maketitle
	\begin{abstract}
		The development of quantum technologies relies on creating and manipulating quantum systems of increasing complexity, with key applications in computation, simulation, and sensing. This poses severe challenges in efficient control, calibration, and validation of quantum states and their dynamics. Machine learning methods have emerged as powerful tools owing to their remarkable capability to learn from data, and thus have been extensively utilized for different quantum tasks. This paper reviews several significant topics related to machine learning-aided quantum estimation and control.  In particular, we discuss neural networks-based learning for quantum state estimation, gradient-based learning for optimal control of quantum systems, evolutionary computation for learning control of quantum systems, machine learning for quantum robust control, and reinforcement learning for quantum control. This review provides a brief background of key concepts recurring across many of these approaches with special emphasis on neural networks, evolutionary computation, and reinforcement learning.
	\end{abstract}
 

\section{Introduction}\label{sec:introduction}

Estimation and control of quantum systems are fundamental in advancing quantum technologies, experiencing notable progress over the past three decades; for an overview, see, e.g., the survey papers~\cite{dong2010quantum,rabitz2000whither,brif2010control,glaser2015training,altafini2012modeling} and monographs~\cite{nielsen2010quantum,alessandro2008introduction}. Acquiring information about unknown quantum entities can be realized by performing measurements on quantum systems and deducing patterns from measured results. Owing to the considerable ability of machine learning (ML) to extract useful patterns in large-scale and complex data, it is highly desirable to apply ML to assist in the post-processing of measurement data. Quantum control, on the other hand, focuses on directing the evolution of quantum systems with the objective often being to maximize a specific performance function~\cite{rabitz2000whither}. ML offers distinct advantages in searching for control policies without knowing the exact model of quantum systems~\cite{biamonte2017quantum}. In this review, we present a comprehensive introduction to both quantum estimation and quantum control tasks, highlighting the integration of ML techniques within these domains~\cite{carleo2019machine}.   

	
Stemming from the traditions of pattern recognition, such as recognizing handwritten text, and statistical learning theory, ML addresses a variety of learning scenarios~\cite{dunjko2018machine}, including learning from data, e.g., supervised (data classification), and unsupervised (data clustering) learning, or learning from interaction, e.g., reinforcement learning  (decision making). One particularly attractive model in ML is artificial neural networks (ANNs, or just NNs), where researchers have found multiple, sequential, hidden feedforward layers, i.e., deep NNs, may have additional benefits~\cite{goodfellow2016deep}. For example, convolutional NNs (CNNs), known for their translation invariance, have achieved success in the fields of vision and pattern recognition~\cite{rawat2017deep}. Recurrent neural networks (RNNs) have been proposed for dealing with sequential data or time series data (see~\cite{lipton2015critical} for detailed information), with various models, e.g., long short-term memory (LSTM)~\cite{graves2012long} and gated recurrent unit (GRU)~\cite{chung2014empirical}. Transformers employ an
 attention mechanism to capture long-range dependencies more effectively~\cite{vaswani2017attention}, thereby achieving great success in natural language processing and computer vision~\cite{han2022survey}. In the field of generative models, generative adversarial networks (GANs)~\cite{salimans2016improved} and variational autoencoders (VAs)~\cite{doersch2016tutorial} are two successful approaches. The former involves a generative network and a discrimination network, while the latter involves an encoder and a decoder, where each part can be built using different NN architectures. Diffusion models operate by first transforming images into Gaussian distributions through a forward diffusion process, then iteratively sampling the new images from this noisy state using a reverse denoising process, showcasing exceptional capabilities in image synthesis~\cite{ho2020denoising,nichol2021improved,dhariwal2021diffusion}.
Another type of ML is reinforcement learning (RL) (see~\cite{sutton1999reinforcement} for a review), which was initially developed for robotics but has been extensively applied to other fields that involve sequential decision-making processes (e.g., Alphago~\cite{silver2016mastering}). {\color{black}Recently, quantum-mechanical formalism has been incorporated into ML, known as quantum machine learning, and has demonstrated ``quantum advantage" in sample complexity or time complexity when dealing with data originating from quantum systems,  see~\cite{martin2022quantum,cerezo2022challenges} for a review}. 


{\color{black}It is a fundamental task to characterize the state or the evolution of a quantum system. This typically involves reconstructing full or partial characteristics from measured statistics, collectively referred to as, learning a complex distribution. ML provides a data-driven technique to extract useful patterns from data, which suggests a natural benefit of robustness against noise in measurement data~\cite{palmieri2020experimental}.} For example, NNs have demonstrated an intrinsic capability of efficiently representing quantum properties in a generative learning way~\cite{carleo2017solving,carleo2019machine,carrasquilla2021use}. Within the framework of function approximation by learning from labeled data, NNs have been widely investigated for quantum state tomography (QST)~\cite{lohani2020machine,neugebauer2020neural,schmale2022efficient,wang2022ultrafast,lohani2023dimension}. Among them, different architectures have been applied, with the Transformer architecture being used to capture long correlations among constituent qubits, i.e., quantum entanglement~\cite{cha2021attention}. Compared to conventional methods, NNs aim to capture key patterns by approximating a complex function from large-scale data. This ability brings robustness against possible noise in the data, making NNs promising for reconstructing quantum states from imperfect measurement data~\cite{ma2024neural,palmieri2020experimental}. Inspired by the remarkable expressivity of NNs, there are also efforts to build a variational ansätze using quantum circuits (also called quantum NNs) (see~\cite{liao2024quantum,nguyen2024theory} for a review). Through deliberate designs, these components can be leveraged for tasks such as estimating wavefunctions~\cite{liu2020variational} and reconstructing unitary operations~\cite{xue2022variational}. Drawing from classical autoencoders~\cite{doersch2016tutorial}, quantum autoencoders have been proposed to reorganize high-dimensional states into latent representations that can be potentially recovered with high fidelity, thus saving valuable resources~\cite{wan2017quantum,romero2017quantum,huang2020realization,ma2023compression,ma2024quantum}. Additionally, quantum metrology studies the estimation of the parameters of quantum systems, {\color{black}which relies on identifying optimal probe states, evolution processes, and measurement operators~\cite{giovannetti2011advances}. ML methods offer a distinct solution to adaptive learning of quantum systems. For example, an adaptive Bayesian approach updated the evolution time, contributing to the efficient use of resources (i.e., the number of experiments) for phase estimation~\cite{fiderer2021neural}}. 



Another significant task in quantum technology is the design of a target quantum evolution, which can be tackled by quantum control. It's goal is to identify how the control fields of physical systems can be adapted to achieve the desired evolution~\cite{dong2010quantum}. 
This underlying problem often manifests as an optimization problem under realistic constraints, posing challenges for conventional optimizers. Learning-based control approaches have been developed for the manipulation of various quantum systems~\cite{rabitz2000whither}, where different learning algorithms (e.g., greedy algorithms~\cite{khaneja2005optimal,Chen2014PRA} or global approaches~\cite{zahedinejad2014evolutionary,zahedinejad2015high,zahedinejad2016designing,ma2015differential}) iteratively suggests improved control fields based on prior trial experiments~\cite{rabitz2000whither,chen2013closed,dong2020learning}. 
By incorporating the concept of sampled-based learning, the optimized control pulses exhibit robustness against uncertain parameters in system Hamiltonians~\cite{dong2020learning}. Complementary to learning-based optimization, identifying optimal strategies can also be realized with real-time feedback from quantum systems~\cite{sivak2023real,reuer2023realizing}. 
This constitutes an active learning process where an RL \emph{agent} is designed to learn a policy rather than the optimization of a particular control field~\cite{niu2019universal,fosel2018reinforcement,ma2022curriculum}. This model-free approach allows for more autonomy and flexibility (i.e., the same machinery can be used in additional settings without alteration). Incorporating NNs into RL not only enables flexible representations of a \emph{state} (e.g., wave function, density matrix) and an \emph{action} (e.g., discrete or continuous controls) but also makes it possible to learn a robust control policy by learning from large-scale data~\cite{niu2019universal,jiang2022robust}. Flexible representation using NNs accommodates the inherent properties of quantum stochasticity and partial observability. This is significant for quantum experiments when only partial observations of quantum systems are available (see e.g., \cite{sivak2022model,jiang2022robust,fosel2018reinforcement}). Deep reinforcement learning (DRL) methods have been extensively applied to quantum error correction~\cite{fosel2018reinforcement,sivak2022model,sivak2023real} and other applications (see~\cite{chen2024efficient,moro2021quantum} for quantum compiling, and ~\cite{xu2019generalizable,xu2021generalizable,qiu2022efficient,fallani2022learning} for quantum metrology).

In this review, we attempt to provide a selected overview of ML's diverse applications in quantum technologies.
Specifically, we delve into quantum estimation challenges where ML can aid by leveraging data-driven learning techniques. Additionally, we address the complexities involved in controlling quantum systems by utilizing ML methods. The remainder of this paper is organized as follows. We provide background information on quantum estimation, quantum control, and several fundamental concepts in quantum mechanics in Section~\ref{Sec:pre}. Section~\ref{Sec:quantumestimate} delves into the accomplishments achieved through the integration of ML in quantum estimation tasks. Section~\ref{Sec:learningcontrol} investigates the performance of learning-based optimization of quantum systems and Section~\ref{Sec:feedback} focuses on the utilization of RL for control of quantum systems. Finally, we conclude with an outlook in Section~\ref{Sec:conclusion}.

\section{Preliminaries}\label{Sec:pre}
	
In this section, we briefly introduce several related concepts for estimation and control of quantum systems including quantum states, quantum measurements, and quantum evolution. Then, we introduce several concepts related to machine learning methods. 

 
	\subsection{Fundamental concepts in quantum mechanics}
	\textbf{Quantum state.}
	In quantum mechanics, the state of a finite-dimensional closed quantum system can be represented by a unit complex vector $|{\psi}\rangle$. This notation is known as the Dirac representation~\cite{dirac1981principles}. The state $|\psi\rangle$ is also called a wavefunction in a complex Hilbert space $\mathbb{H}$, which is useful in describing a pure state. Information can be encoded using two-state (two-level) quantum systems (called qubits) and the state $|\psi\rangle$ of a qubit can be written as 
	\begin{equation}
		|\psi\rangle = a_0 |0\rangle + a_1 |1\rangle
	\end{equation}
	where $a_0,a_1\in \mathbb{C}$ and $|a_0|^2+|a_1|^2=1$. Here, $|0\rangle$ and $|1\rangle$ correspond to the states zero and one for a classical bit~\cite{nielsen2010quantum}. Since the global phase of a quantum state has no observable physical effect, we often say that the vectors $|\psi\rangle$ and $e^{\mathrm{i}\phi}|\psi\rangle$ (where $\mathrm{i}=\sqrt{-1}$ and $\phi\in \mathbb{R}$) describe the same physical state. 
	
	In practical applications, quantum systems are usually not closed. They may be open quantum systems and their states cannot be written in the form of unit vectors~\cite{altafini2004coherent}. In such a case, the density operator $\rho$ is defined to describe the states of open quantum systems. Let $(\cdot)^{\dagger}$ denote the adjoint operation. A density operator $\rho$ is Hermitian and positive semidefinite, and of trace one, i.e., satisfying $\rho=\rho^{\dagger}$, $\rho \geq 0$, $\textup{Tr}(\rho)=1$. 
 A density operator can be represented as an ensemble of pure states $\{|\psi_j\rangle\}$, i.e., 
	\begin{equation}
		\rho = \sum_j p_j |\psi_j\rangle \langle \psi_j|,
	\end{equation}
	where $\langle \psi_j| =(|\psi_j\rangle)^{\dagger}$, $p_j  \geq 0$ and $\sum_j p_j=1$.  For a pure state $|\psi\rangle$, we have $\rho=|\psi\rangle \langle \psi|$ and $\textup{Tr}(\rho^2)=1$. 

	\textbf{Quantum measurement.}
	In quantum control and engineering, it is important to extract information from quantum systems. Measurement theory in quantum mechanics
	is essentially different from that in classical physics since
	a measurement on a quantum system unavoidably affects the
	measured system (a detailed discussion of this issue can be
	found in~\cite{breuer2002theory}). 
	A quantum measurement is associated with a collection $\{\mathcal{M}_j\}$ of measurement operators, acting on the state space of the system being measured and satisfying the completeness equation
	\begin{equation}
		\sum_j \mathcal{M}_j^{\dagger} \mathcal{M}_j = \mathcal{I}, 
	\end{equation}
 where $ \mathcal{I}$ denotes the identity matrix and the index $j$ labels the possible measurement outcomes. For a quantum system in the state $|\psi\rangle$, the probability that the $j$-th result occurs is given by
	\begin{equation}
		p_j = \langle \psi| \mathcal{M}_j^{\dagger} \mathcal{M}_j |\psi\rangle.
		\label{eq:collapsepure}
	\end{equation}
	If we obtain the outcome $j$, the state of the measured system changes to $\mathcal{M}_j |\psi\rangle / \sqrt{p_j}$. The completeness equation is equivalent to requiring that all the probabilities sum to one, i.e., $\sum_j p_j=1$. 
 {\color{black}When the post-measurement state of the quantum system
is of no interest, a measurement for a quantum system can be
characterized using the positive operator-valued measure (POVM). Specifically, a set of operators $E_j$ are known as the POVM elements associated with the measurement, and the corresponding probability is given by $p_j=\langle \psi| E_j | \psi\rangle$. A widely used measurement model is projective measurement, satisfying $\mathcal{M}_j=\mathcal{M}_j^{\dagger}, \mathcal{M}_{j}\mathcal{M}_{j^{\prime}}= \delta_{jj^{\prime}} \mathcal{M}_j$, where $\delta_{jj^{\prime}}$ represents the Kronecker delta. Define projectors as $\mathcal{P}_j = \mathcal{M}_j^{\dagger} \mathcal{M}_j$ and the probability of the $j$-th outcome is given as $p_j = \langle \psi| \mathcal{P}_j |\psi\rangle$. }

	\textbf{Quantum evolution. }
	The evolution of a closed quantum system can be described by a unitary transformation. That is, the state $|\psi\rangle$ of the system at time $t_1$ is related to the state $|\psi^{\prime}\rangle$ of the system at time $t_2$ by a unitary operator $\mathcal{U}$ as 
	\begin{equation}
		|\psi^{\prime}\rangle = \mathcal{U} |\psi\rangle,
	\end{equation}
	where $\mathcal{U} \mathcal{U}^{\dagger}= \mathcal{I}$. For mixed states, we have $\rho^{\prime} = \mathcal{U} \rho \mathcal{U}^{\dagger}$. 
	Quantum gates can be expressed as unitary operators. For example, the \emph{Hadamard} gate has the corresponding unitary matrix 
	$$\frac{1}{\sqrt{2}}\left[\begin{array}{cc}
		1 & 1 \\
		1 & -1
	\end{array}\right].$$
	
	If the system under consideration has interaction with its environment, it becomes an open quantum system and has a more complicated state evolution. To deal with this situation, quantum processes are proposed to describe the time evolution of an (open) quantum system (also known as a quantum dynamical map), which is a linear map from the set of density matrices to itself. Let $\Lambda$ be a map that transforms an input state $\rho_{\text{in}}$  into an output state 
	\begin{equation}
		\rho_{\text{out}} = \Lambda(\rho_{\text{in}}). 
	\end{equation}
 For a physical quantum map, $\Lambda$ must be completely positive (CP). 

 According to the Choi-Jamiolkowski isomorphism~\cite{choi1975completely}, there exists a one-to-one correspondence between every quantum map $\Lambda$ and a Choi operator $Q_{\text{Choi}}$ ~\cite{choi1975completely}, such that
\begin{equation} 
	\Lambda(\rho_{\text{in}})=\textup{Tr}_A(Q_{\textup{Choi}}(\rho_{\text{in}}^T\otimes  \mathcal{I})),
		\label{eq:choi}
	\end{equation}
	where $\textup{Tr}_A(\rho)$ denotes the partial trace corresponding to the subsystem  $A$~\cite{nielsen2010quantum}, and we have
	\begin{equation}
		Q_{\textup{Choi}}=\sum_{ij}  |i\rangle \langle j| \otimes  \Lambda(|i\rangle \langle j|),
		\label{eq:choi}
	\end{equation}
	indicating that $Q_{\textup{Choi}}$ characterizes $\Lambda$ completely. 
 


 
	\subsection{Estimation of quantum systems}
In most tasks in quantum information and quantum engineering, it is required to obtain enough information about the unknown quantum entity; i.e., information about certain key structures or parameters of the entity needs to be extracted. This highlights the significance of quantum estimation, which is often called quantum tomography (QT). Distinct from the classical information, QT usually assumes a framework where a large number of independent identical copies of an unknown quantum state are available, and data are obtained through proper interaction (e.g., quantum measurement) with these copies following certain protocols. {\color{black}To uniquely determine a quantum state, a set of informationally complete (or overcomplete) measurements are performed, with measured statistics given as 
$\boldsymbol{f}=[f_1,f_2,...]^T$, with $\sum_i f_i  = 1$}.

QT relies on the measured frequency vector $\boldsymbol{f}=[f_1,f_2,...]^T$ which is a statistical approximation to the true probability vector $\boldsymbol{p}=[p_1,p_2,...]^T$ (see Eq.~(\ref{eq:collapsepure}) for more details), in order to infer underlying information about quantum entities. Such a task can be summarized as obtaining an estimate of the entire entity (called full QT) or of partial properties of the entity. Following this framework, the estimation of quantum states is realized by determining the density matrices of a fixed state of quantum systems, while the estimation of the quantum process is realized by determining the evolution of quantum systems.
	
	\subsection{Quantum control systems}
	
	For a closed quantum system, its dynamics can be described by the Schr\"{o}dinger equation:
	\begin{equation}
		\frac{\mathrm{d}}{\mathrm{d}t}|{\psi}(t)\rangle=-\frac{\rm{i}}{\hbar}\mathcal{H}(t) |{\psi(t)}\rangle,
		\label{eq:schron}
	\end{equation}
	where $\hbar$ is the Planck constant and hereafter we use atomic units to set $\hbar=1$. $\mathcal{H}(t)$ is a Hermitian operator known as the Hamiltonian of the quantum system. It also has a density matrix version, which is the Liouville-von Neumann equation
	\begin{equation}
		\dot{\rho}(t)=-\mathrm{i}[\mathcal{H}(t), \rho(t)],
	\end{equation}
	where $[A,B]=AB-BA$ is the commutator.  For a quantum control system, we may consider its system Hamiltonian as follows
	\begin{equation}
		\mathcal{H}(t)=\mathcal{H}_0+\sum_{m=1}^{N_c}u_m(t)\mathcal{H}_m,
  \label{eq:Hamiltonian}
	\end{equation}
	where $\mathcal{H}_0$ denotes the time-independent free Hamiltonian of the system, and the control Hamiltonian operators $\mathcal{H}_m$ represent the interaction of the system with the control fields. The unitary evolution $\mathcal{U}(t,t_0)$ from time $t_0$ to $t$ under the Hamiltonian can be given as 
 \begin{equation}
\mathcal{U}(t,t_0)=\mathcal{T}_{\leftarrow}[\exp(-{\rm{i}}\int_{t_0}^{t}\mathcal{H}(t^{\prime})dt^{\prime})],
\label{eq:unitary}
\end{equation}
where $\mathcal{T}_{\leftarrow}$ represents time-ordering~\cite{breuer2002theory}. 
Quantum control aims at searching for a set of control fields $\{u_m(t)\}$ to drive the quantum system to achieve a given target with desired performance.

	When a quantum system interacts with its environment (i.e., a dissipative bath coupled to a quantum system), the system becomes an open one and its dynamics under Markovian approximation can be described by the Markovian master equation (MME)~\cite{dong2010quantum}:
	\begin{equation}
		\dot{\rho}(t)=-{\rm{i}}[\mathcal{H}(t),\rho(t)]+\sum_k \eta_k \mathcal{D}[L_k](\rho(t)),
		\label{eq:lindblad}
	\end{equation}
	with $\mathcal{D}[L_k](\rho)=L_k \rho L_k^{\dagger}-\frac{1}{2} L_k^{\dagger} L_k \rho-\frac{1}{2}\rho L_k^{\dagger} L_k$, where $\{L_k\}$ are the operators coupling with the environment and the coefficients $\eta_k\geq 0$ characterize the relaxation rates. 
 
 In feedback control, it is required to continuously monitor a quantum system to obtain feedback information. {\color{black}The evolution of a quantum system under continuous homodyne measurements of a field observable coupled with the system through an operator $L$ can be described by the following stochastic master equation (SME):
 \begin{equation}
 \begin{aligned}
		& \dot{\rho}(t) =-\mathrm{i}[\mathcal{H}(t), \rho(t)] +\kappa \mathcal{D}[L](\rho(t))+ \sqrt{\kappa} \mathscr{H}[L](\rho(t)) \mathrm{d} W_t,\\&
  \mathscr{H}[L](\rho(t)) = L\rho(t)+\rho(t)L^{\dagger}-\langle L+L^{\dagger}\rangle \rho(t),
  \label{eq:sme}
  \end{aligned}
	\end{equation}
 where $\kappa \in (0,1]$ is a parameter related to the measurement strength and $\langle \cdot \rangle=\textup{Tr}(\cdot  \rho)$. $\mathrm{d} W_t$ is a Wiener increment with zero mean and variance equal to $\mathrm{d}t$ and satisfies the following relationship to the measurement output $y_t$:
	\begin{equation}
	 \mathrm{d} y_t=\mathrm{d} W_t+ \sqrt{\kappa} \langle L+L^{\dagger} \rangle \mathrm{d} t.
		\label{eq:wiener}
	\end{equation}
Note that Eq.~(\ref{eq:sme}) is only a typical form of SME and there exist many different types of SMEs that depend on different measurement processes~\cite{qi2009quantum}.}

 \subsection{Machine learning methods}\label{sec:ML}

{\color{black}ML is a branch of artificial intelligence that focuses on developing and studying statistical algorithms capable of learning from data and generalizing to unseen data, enabling systems to perform tasks without explicit instructions}. Let us define it as an \emph {agent}. Central to this approach is the availability of large amounts of data (or the possibility of synthetically generating it). The way that the \emph{agent} is trained depends on the given task and is generally divided into the following categories.
	\begin{itemize}
		\item Supervised learning. The training data are labeled with their target values: that is, the labels that should be learned by the \emph{agent} are known for the training data.
		\item Unsupervised learning. The training data are not labeled, and the \emph{agent} is trained to recognize the structure or pattern in the data.
		\item Reinforcement learning. There is no training data required, while the \emph{agent} interacts with an \emph{environment}\footnote{The \emph{environment} in RL contexts is distinguished from the environment (or a dissipative bath that is coupled to a quantum system) in quantum contexts.} Their interaction generates data that can be used for training the \emph{agent} to maximize a \emph{reward} assigned according to the \emph{agent}’s \emph{action}. 
	\end{itemize}
ML can be used to address various tasks that can be grouped into different types. For instance, typical ML tasks can be divided into the classification of data into categories, the regression of functions given their values on data vectors, and the sampling of new data vectors that have a similar distribution to vectors in the given data, which is also called a generative model.

The basic building block of modern ML architectures can be expressed as an artificial neuron. Its basic units are single-output nonlinear functions: $y=g(W x+b)$, where $g: \mathbb{R}^n \rightarrow \mathbb{R}$ is a nonlinear activation function, and the weights $W$ and optional-biases $b$ are parameters to be optimized during the training phase. As a single neuron is not sufficient to approximate complex dependencies, multiple neurons are arranged and connected to form a multilayer NN. Generally, NNs with at least a single hidden layer can approximate arbitrary functions (the NNs are usually very wide for complex functions), which forms the theoretical basis for using them in approximating relationships between different types of data~\cite{goodfellow2016deep}. {\color{black}A fully connected multilayer NN is called a multilayer perceptron (MLP). Different from MLPs that utilize fixed activation functions (i.e., fixed form of $g$) on nodes (``neurons"), Kolmogorov–Arnold Networks (KANs), featuring learnable activation functions, have emerged as a promising tool in ML community~\cite{liu2024kan}.}
To train NNs, one needs to choose a problem-specific cost function (e.g., a mean squared error for regression problems or a cross-entropy loss for classification problems) that may be minimized via stochastic gradient descent. A central goal of ML algorithms is generalizability: to achieve the given task for both the given training data and when new (testing) data are provided after the training stage. A large enough NN is known to be a universal function approximator. However, the size of the NN should be carefully chosen as its trainability can be compromised for too large a size, and its generalizability may also decrease in the presence of a high expressivity and long training schedules, which is called overfitting in the ML community. {\color{black}Notably, as the model size increases, performance first gets worse and then gets better, which is called the ``double-descent" phenomenon~\cite{nakkiran2021deep}}.

 
In the RL paradigm, the interaction of the \emph{agent} with its \emph{environment} is usually described within the framework of Markov decision processes (MDPs), defined by a 5-tuple $\langle \mathbb{S},\mathbb{A},\mathbb{P},\mathbb{R},\gamma \rangle$~\cite{sutton2018reinforcement}, where $\mathbb{S}$ denotes a set of \emph{states} that can be observed from the \emph{environment}, $\mathbb{A}$ represents a set of \emph{ \emph{actions}} that can be executed in the \emph{environment}, $\mathbb{P}$ represents the \emph{state} transition probability, $\mathbb{R}$ represents the \emph{reward}\footnote{The \emph{state} in RL contexts is distinguished from the state in quantum contexts. To emphasize other key elements in the RL community and maintain consistency, we also use \emph{action} and  \emph{reward} in this paper.}, $\gamma\in[0,1]$ is the discount factor. A policy $\pi$ is defined as a mapping from the state space $\mathbb{S}$ to the \emph{action} space $\mathbb{A}$, i.e., $\pi: \mathbb{S} \rightarrow \mathbb{A}$. RL aims at determining an optimal  \emph{action} $a_t^{*}$ at each state $s_{t}$ to maximize the cumulative discounted future \emph{rewards} $G_t =
\sum_{k=0}^{T-t}\gamma^{k}r_{t+k}$. To this end,  the \emph{reward} signal is designed by a human supervisor to indicate how good the new \emph{state} is after the applied \emph{action}. Importantly, it is possible to specify the \emph{reward} signal for achieving a final goal without knowing the optimal \emph{action}, which is a major difference between RL and supervised learning. In the RL community, the \emph{agent} interacts with its \emph{environment} whose \emph{state} can be either fully or only partially observed through a corresponding observation obtained after executing an \emph{action} according to an underlying policy $\pi$. In this case, partially observable MDPs are proposed, where the observation is dependent on the current \emph{state} and the previous \emph{action}s~\cite{kaelbling1998planning}. 
RL methods can be classified into three categories in Table~\ref{table:RL}: (i) value-based methods that first approximate value functions, e.g., $Q(s, a)$ represents the expected cumulative \emph{reward} after taking \emph{action} $a$ in \emph{state} $s$~\cite{watkins1992q}) and then obtain a policy, e.g., $a^*=\max_{a \in \mathbb{A}} Q(s,a)$; (ii) policy-based methods that directly approximate a policy function $a=\pi(s)$~\cite{sutton1999policy}; (iii) actor-critic methods that combine value approximation and policy approximation. Notably, by approximating the value function or policy function using multilayer NNs, deep RL methods represent a step toward building autonomous systems that can accept raw data from the real world~\cite{arulkumaran2017deep,mnih2015human}, without relying on (manually) designed feature vectors.


\begin{table}
		\centering
		\caption{A taxonomy of RL. {\color{black} Different methods can be classified into (i) value-based methods that optimize value functions; (ii) policy-based methods that optimize policy functions; and (iii) actor-critic methods that jointly optimize value functions and policy functions.}}
		\begin{tabular}{cc}
			\hline
			\textbf{(i) Value-based algorithms }                                              & \textbf{(ii) Policy-based algorithms}                                  \\ \hline
			Q-learning\cite{watkins1992q}   &  Policy gradient\cite{sutton1999policy}                                                                                     \\
			SARSA\cite{rummery1994line}    &           Trust region policy optimization \cite{schulman2015trust}                     \\
			Deep Q network (DQN)\cite{mnih2013playing} & Proximal policy optimization (PPO)\cite{schulman2017proximal}
			\\ \hline
			\multicolumn{2}{c}{\textbf{(iii) Actor-critic algorithms}: learn policy and value functions jointly} 
			\\ \hline
			\multicolumn{2}{c}{Asynchronous advantage actor-critic (A3C)~\cite{mnih2016asynchronous}}                                                       \\
			\multicolumn{2}{c}{Deep deterministic policy gradient (DDPG)~\cite{lillicrap2015continuous}}                                                     \\
			\multicolumn{2}{c}{Twin Delayed DDPG (TD3)~\cite{fujimoto2018addressing}}  
			\\ \hline                                                              
		\end{tabular}
		\label{table:RL}
	\end{table}

\section{Machine learning for quantum estimation}\label{Sec:quantumestimate}

Quantum estimation usually involves the reconstruction of full or partial characteristics from measured statistics, whose performance may be limited by the state-preparation-and-measurement (SPAM) errors. Machine learning provides a means of building noise resilience into the post-processing of measurement data and, thus can be useful to assist in quantum estimation tasks. In the following, we first outline the process of converting quantum estimation into an inversion problem in Subsection~\ref{Subsec:inversion}. Subsequently, we focus on QST and investigate the performance of machine learning for quantum estimation in Subsection~\ref{SubSec:QST}. Then, we present results on the estimation of quantum dynamics in Subsection~\ref{SubSec:QPT}. This section concludes with a discussion of the outlook and open questions of ML-aided quantum estimation in Subsection~\ref{Subsec:Summary}.

\subsection{Quantum estimation as an inversion task}\label{Subsec:inversion}
	
\begin{figure}
    \centering
     \includegraphics[width=0.5\textwidth]{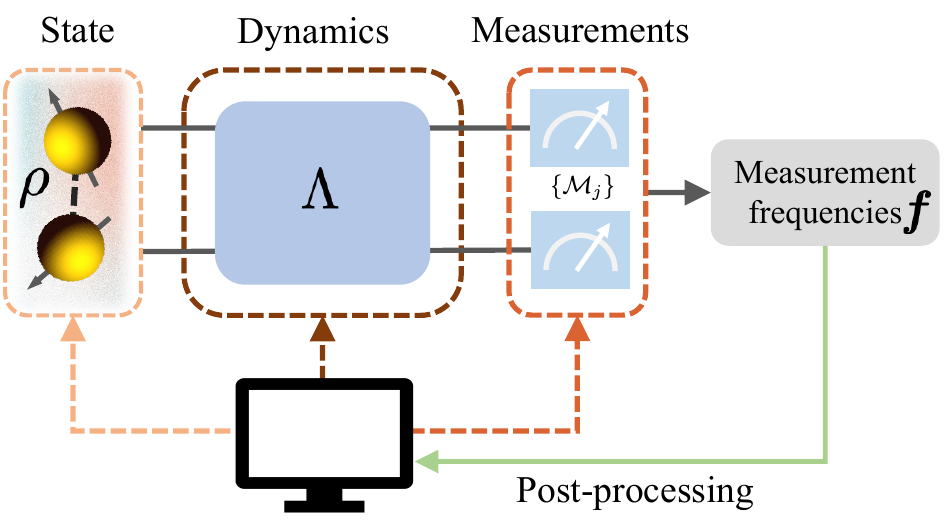}
\caption{Schematic of quantum estimation, including estimating quantum states, dynamics, and measurements. An initial quantum state $\rho$ undergoes a quantum operation $\Lambda$, ending up with an output state. Measurement frequencies $\boldsymbol{f}$ are collected by performing quantum measurements $\{\mathcal{M}_j\}$ on the output state. Quantum estimation aims to capture the underlying pattern among the observed data, which can be individually applied to deduce the parameters for the quantum state $\rho$, quantum evolution $\Lambda$, and quantum measurements $\{\mathcal{M}_j\}$, respectively. }
		\label{fig:QT}
	\end{figure}

Before introducing different solutions to quantum estimation, we first provide a broad overview of learning approaches for quantum systems at different levels involving quantum states, quantum dynamics, and quantum measurements using post-processing techniques (see Fig.~\ref{fig:QT}). Although the characterization of quantum systems can be realized using traditional methods (such as linear regression estimation~\cite{qi2013quantum}, maximum likelihood estimation~\cite{jevzek2003quantum}, and Bayesian estimation~\cite{blume2010optimal}), their solutions usually rely on informationally complete measurements or a large number of measurement copies. The complexity of quantum systems scales exponentially with their size, but in many practical scenarios, certain assumptions like low rank, sparsity, or specific dynamics, make it possible for classical algorithms to efficiently characterize unknown quantum entities~\cite{gebhart2023learning}. 

For quantum estimation, the general procedure is to collect measured data for the estimation of parameters of quantum systems, which can be regarded as an inversion problem. The emergence of ML offers an alternative automated procedure to capture the characteristics of quantum entities that match the observed data, i.e., to learn a parameterized function by fitting data, which functions as variational ansätze for quantum systems~\cite{carleo2017solving}. One useful choice for the family of parameterized functions can be NNs, which serve as universal function approximators capable of acquiring mappings from noisy input data to output labels. The introduction of NNs enables the average reconstruction fidelity to be improved between 10\% and 27\% on 2-qubit systems compared to a protocol treating SPAM errors by process tomography and a SPAM-agnostic protocol, respectively~\cite{palmieri2020experimental}.



Inverse problems deal with determining parameters of interest, $\boldsymbol{w} \in \mathbb{W}$, in a problem involving data $\boldsymbol{f} \in \mathbb{F}$. For quantum estimation problems, quantum measurement involving a set of measurement operators $\boldsymbol{\mathcal{M}}=\{\mathcal{M}_i\}$ maps the quantum entity $\boldsymbol{w} $ to measured frequencies $\boldsymbol{f}$ in a forward way. The goal of quantum estimation is to find the inverse of this process as $\boldsymbol{w} = \mathcal{G}(\boldsymbol{f})$, where $\mathcal{G}$ represents a mapping that transforms $\boldsymbol{f}$ into $\boldsymbol{w}$. 
Such problems frequently face the challenge of being ill-posed, especially when noise becomes a primary contributor and can be amplified during the inversion process. 
Additionally, the selection of inappropriate measurement operators or insufficient resources for the measurement process can further exacerbate the ill-posed nature of quantum estimation. While altering the measurement operator or acquiring additional data are straightforward solutions to address these issues, they might not be effective in cases where noise amplification during inversion is excessively high. Hence, it is desirable to consider the continuous dependence of the solution on data and the robustness of the model under noise or perturbations~\cite{bal2012introduction}. 

According to the universal approximation theorem,  deep NNs with multiple fully connected layers can act as universal function approximations from $\mathbb{R}^n \rightarrow \mathbb{R}^m$~\cite{lecun2015deep}. Therefore, they can be used to replace the unknown forward and inverse models and extract information from data. For example,  one can approximate a map $\boldsymbol{w} = \mathcal{G}_{\xi}(\boldsymbol{f})$ to capture the underlying relationships between $\mathbb{W}$ and $\mathbb{F}$. $\mathcal{G}_{\xi}$ is typically a learnable function with parameters $\xi$ to be optimized via minimizing a loss metric that quantifies the disparity between the predictions generated by the NN and the expected labels. The addition of priors to the NN architecture can enable the use of the universal approximation capacity of NNs as well as leverage human knowledge~\cite{raissi2019physics,beucler2021enforcing}. Striking examples include the design of CNNs from a human visual cortex~\cite{lindsay2021convolutional} and the design of Transformers in language models~\cite{vaswani2017attention}. These approaches have been applied in different quantum estimation tasks on 2-4 qubit systems~\cite{palmieri2020experimental,neugebauer2020neural,lohani2021experimental}. 
	

\subsection{ML-based quantum state estimation}\label{SubSec:QST}
Compared to traditional methods, NN-based approaches typically involve optimizing a function that fits a large number of data items with minimal errors. A data-driven approach enables the capture of key patterns from data, thus exhibiting robustness against imperfect data. These two features make NNs suitable for state estimation, which involves large-scale measurement data, and imperfections and errors in the measurement process~\cite{ma2024neural,palmieri2020experimental}. Here, we focus on QST for finite-dimensional discrete-variable systems, such as multi-qubit systems. Many possible architectures of NNs can be employed to characterize quantum states. 

	
\begin{figure*}
		\centering
		\includegraphics[width=0.9\textwidth]{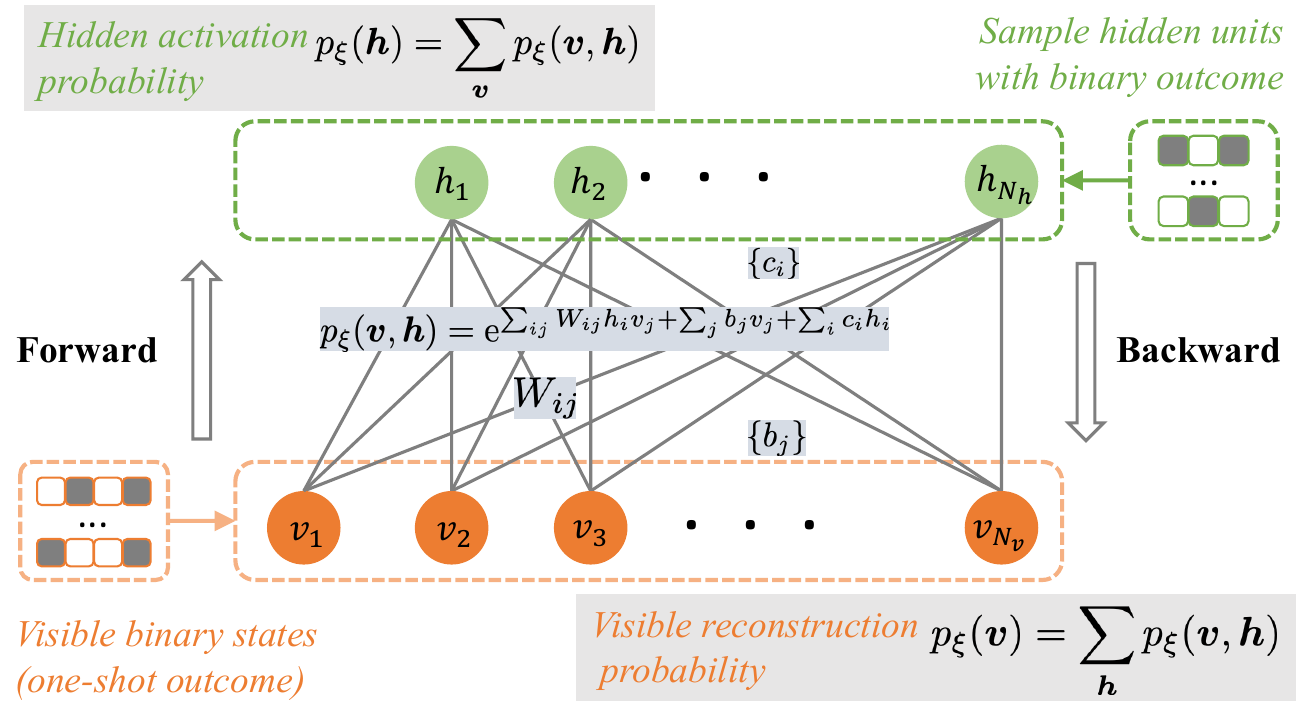}
		\caption{{\color{black}An RBM architecture with 
a parameter vector $\xi$ (corresponding to an amplitude RBM with $\xi=\lambda$ and a phase RBM with $\xi=\mu$). Each RBM features a set of $N_v$ visible neurons (orange dots) and a set of $N_h$ hidden neurons (green dots) and $\xi$ consists of the weights $W$ connecting the layers, and the biases $b$ and $c$ coupled to visible and hidden neurons, respectively.  A  Gibbs distribution (with normalization omitted) is obtained via $ p_{\xi}(\boldsymbol{v}, \boldsymbol{h})=\mathrm{e}^{\sum_{i j} W_{i j} h_i v_j+\sum_j b_j v_j+\sum_i c_i  h_i}$ and the distribution over the visible (hidden) layer is obtained by marginalization over the hidden (visible) degrees of freedom~\cite{goodfellow2016deep}.  In the forward process, visible binary outcomes, i.e., one-shot measurement outcomes, are injected into the RBM to compute the hidden activation probability as $p_{\xi}(\boldsymbol{h})=\sum_{\boldsymbol{v}}p_{\xi}(\boldsymbol{v},\boldsymbol{h})$. In the backward process, based on the hidden activation probability $p_{\xi}(\boldsymbol{h})$, the hidden units are sampled with binary outcomes of $0/1$ and injected into the RBM to obtain the visible reconstruction probability as $ p_{\xi}(\boldsymbol{v})=\sum_{\boldsymbol{h}} p_\xi(\boldsymbol{v}, \boldsymbol{h})$. Two RBMs are trained to minimize the difference between the actual wavefunction and the reconstructed wavefunction $\psi_{\lambda,\mu}(\boldsymbol{v})$ (refer to Eq.~(\ref{eq:RBM}) for detailed information).}}
		\label{fig:RBM-QT}
	\end{figure*}
 

 
\subsubsection{Restricted Boltzmann machine}	
A restricted Boltzmann machine (RBM) stands out for its energy-based model, sharing many properties of physical models in statistical mechanics~\cite{carleo2017solving,torlai2018neural,torlai2019integrating,melkani2020eigenstate}. RBM states offer a compact variational representation of many-body quantum states, capable of sustaining non-trivial correlations, such as high entanglement, or topological features. As an example, let us describe spin quantum systems using an RBM (see Fig.~\ref{fig:RBM-QT}), which features a visible layer 
 (describing the physical qubits, denoted as a data vector $\boldsymbol{v}=\{v_1,v_2,...,v_k,...\}$) and a hidden layer (of binary neurons, denoted as a hidden vector  $\boldsymbol{h}=\{h_1,h_2,...,h_k,...\}$), fully connected with weighted edges to the visible layer. For QST tasks, each element of the data vector corresponds to the one-shot measurement results, e.g., $\{1,0\}$. Then, the wavefunction of a quantum state can be approximated as 
	\begin{equation}
		\psi_{\lambda,\mu}(\boldsymbol{v}) = \sqrt{\frac{p_{\lambda}(\boldsymbol{v})}{Z_{\lambda}}} \mathrm{e}^{\mathrm{i} \log \left[p_\mu(\boldsymbol{v})\right] / 2},
		\label{eq:RBM}
	\end{equation}
where $p_\lambda(\boldsymbol{v})$ and $p_{\mu}(\boldsymbol{v})$ represent the approximated amplitude and phase of the state from two RBM networks, and $Z_{\lambda}$ is the normalization constant. {\color{black} $\psi_{\lambda,\mu}(\boldsymbol{v})$ acts as a latent model to approximate the wavefunction $\psi(\boldsymbol{v})$. Note that a complete RBM-based QST approach requires two RBMs, while Fig.~\ref{fig:RBM-QT} provides an illustration of an RBM with a unified parameter vector $\xi$ that consist of the weights connecting the layers, and the biases, and coupled to visible and hidden neurons, respectively. Specifically, $\xi=\lambda$ represents an amplitude RBM and $\xi=\mu$ represents a phase RBM. For QST tasks, $\{\lambda,\mu\}$ are optimized by minimizing the distance between the reconstructed wavefunction $\psi_{\lambda,\mu}(\boldsymbol{v})$ and the real wavefunction $\psi(\boldsymbol{v})$}.

 The strength of these connections, specified by the parameters, encodes conditional dependence among neurons, in turn leading to complex correlations among the data variables. The correlations induced by the hidden units are intrinsically nonlocal in space and are therefore well suited to describe many-body quantum systems~\cite{carleo2017solving}. This approach has been extended to represent density matrices for mixed states through auxiliary degrees of freedom embedded in the latent space of its hidden units together with purification~\cite{torlai2018latent}. Furthermore, continuous versions of RBMs can be established by replacing the binary encoding (in Fig.~\ref{fig:RBM-QT}) with Gaussian distribution~\cite{krizhevsky2009learning}. A distinct advantage of RBMs is their ability to learn directly from raw data, such as experimental snapshots from single measurements. However, this method requires separate training for each new quantum state as insights gained during the training for one particular state cannot be directly transferred to other states~\cite{zhu2022flexible}. These issues have stimulated the investigation of more flexible models that can generalize across multiple quantum states.

	\begin{figure*}
		\centering
		\includegraphics[width=0.9\textwidth]{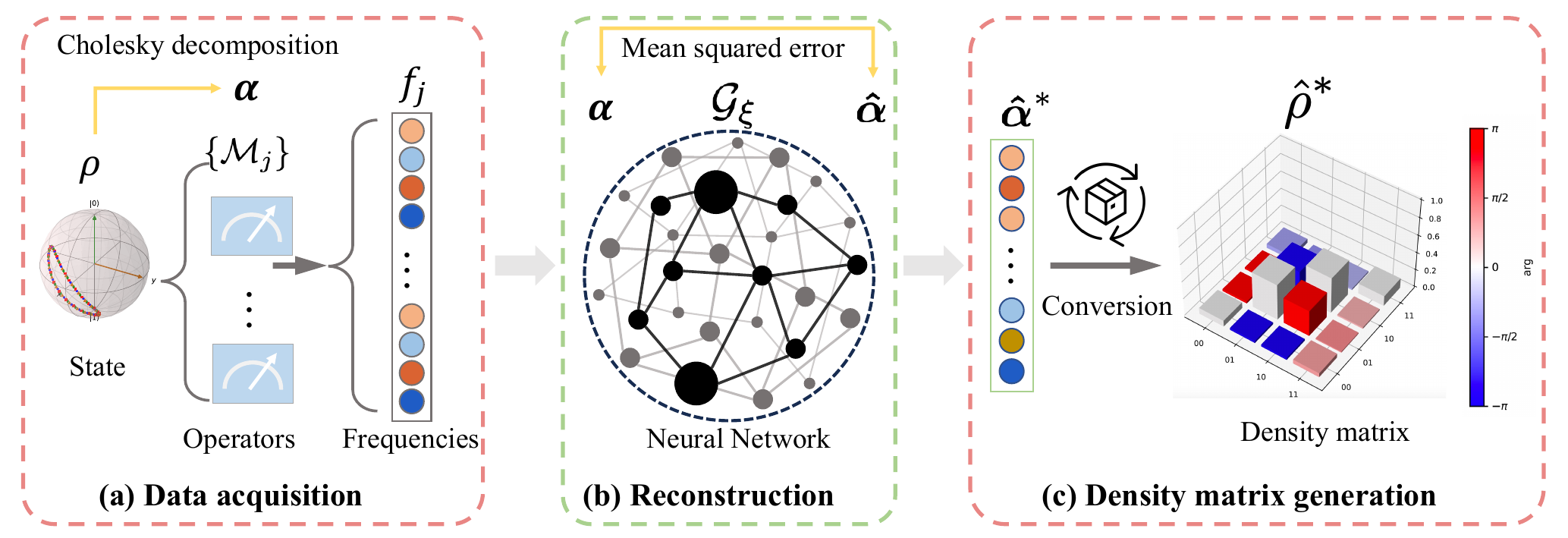}
		\caption{Schematic of NN-based QST. {\color{black}(a) Obtain the measured frequencies $\{f_j\}$ and the ground truth of $\boldsymbol{\alpha}$ via Cholesky decomposition for state $\rho$ ; (b) A multilayer NN model to map the frequency vector $\boldsymbol{f}=[f_1,...f_j,...]^T$ to a predicted real vector $\boldsymbol{\hat{\alpha}}$, and the NN model is trained to minimize the mean squared error between the predicted $\boldsymbol{\hat{\alpha}}$ and the expected value $\boldsymbol{\alpha}$; (c) Obtain a physical density matrix $\hat{\rho}^{*}$ from the optimal $\boldsymbol{\hat{\alpha}}^{*}$. }}
		\label{fig:framework}
	\end{figure*}

	\begin{figure*}
		\centering
		\includegraphics[width=0.9\textwidth]{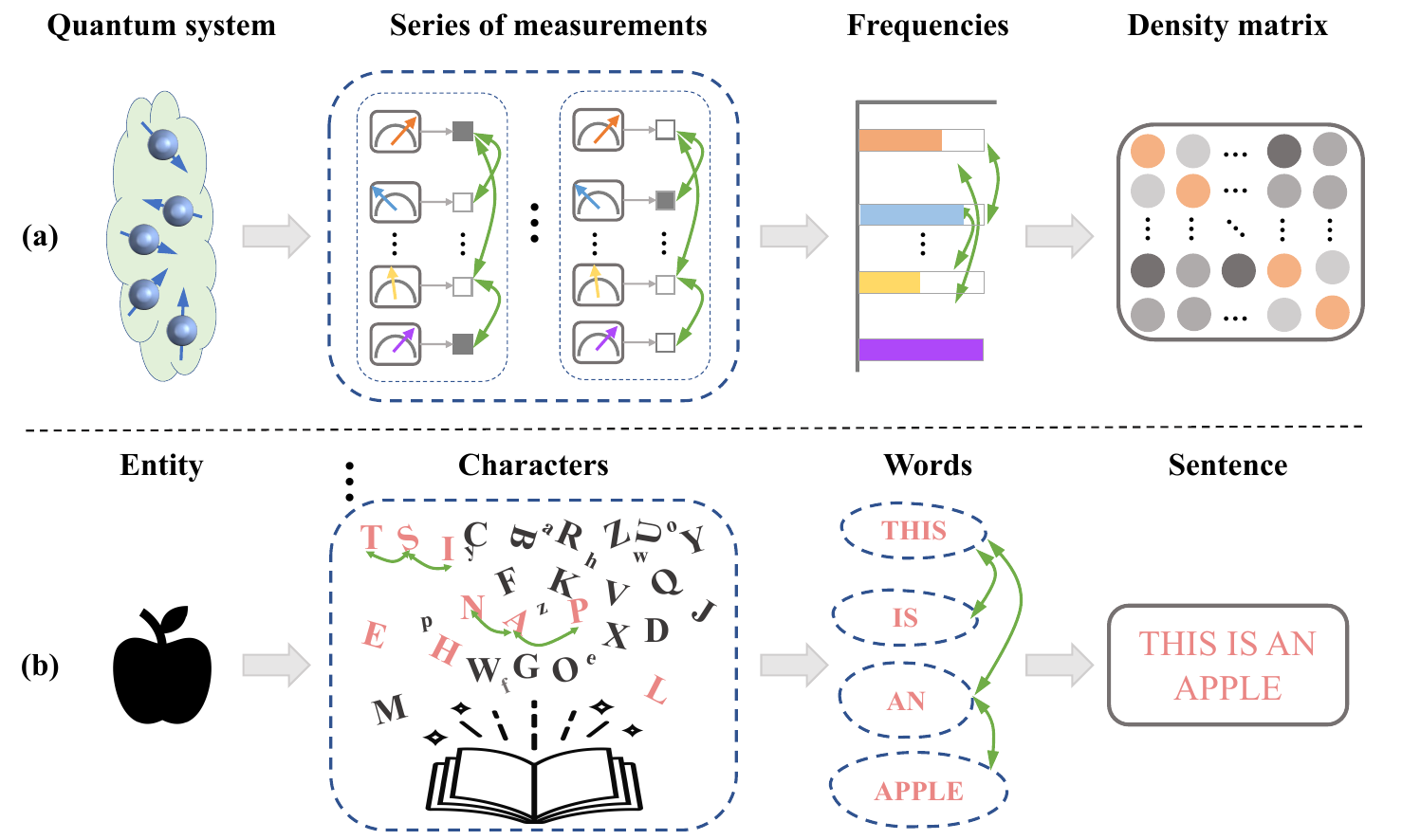}
		\caption{Similarity between QST using structured measurements and the language model using words and characters. (a) Given a quantum system, a series of measurements are performed, each implemented many times, with every one-shot outcome marked as ``0/1". Those outcomes are gathered into frequencies that can be utilized to reconstruct the complete state (density matrix) of the involved quantum system. (b) Given an apple, characters are chosen to specify this object, which can be composed into words. Several words then finally specify a full sentence.}
		\label{fig:language model}
	\end{figure*}

\subsubsection{Feedforward networks}\label{subsec:ffn}
 Feedforward networks are another class of models to approximate a map function from multiple samples, unlike the RBM which focuses on learning a latent model for one quantum state.  Hence, one can build a multilayer network to approximate a function $\boldsymbol{f} \rightarrow \boldsymbol{w}$, where the key is to generate positive semi-definite (PSD) Hermitian matrices from NNs. According to the Cholesky decomposition~\cite{higham1990analysis}, for any Hermitian definite positive matrix $\rho_H=\rho_H^{\dagger} \geq 0$\label{symHermitian}, there exists a lower triangular matrix $\rho_L$\label{symtriangular} such that $ \rho_H =\rho_L \rho_L^{\dagger} $. Conversely, given any lower triangular matrices $\rho_L$, one can obtain a density matrix as:
	\begin{equation}
		\rho = \frac{\rho_L \rho_L^{\dagger}}{\textup{Tr}(\rho_L \rho_L^{\dagger})}.
		\label{eq:tau2rho}
	\end{equation}
This approach can be extended to generate other quantum entities, for example, POVM elements of quantum measurements~\cite{ma2023estimation} and Choi matrices for quantum processes~\cite{ma2023tomography,ahmed2021quantum}, as they involve PSD Hermitian matrices. In particular, POVM elements can be obtained by normalizing a set of lower-triangular matrices $\rho_L^k$
\begin{equation}
 \mathcal{P}_{k} = G^{-1}\rho_L^k (\rho_L^k)^{\dagger} G^{-1}, \quad G = \sqrt{\sum_k \rho_L^k (\rho_L^k)^{\dagger}}.
 \nonumber
\end{equation}
Similarly, Choi matrices can be generated via 
\begin{equation}
\begin{aligned}
    & Q_A = \sqrt{\mathrm{Tr}_{B}(\rho_L\rho_L^{\dagger})},\\ & Q_{\textup{Choi}} = (Q_A^{-1}\otimes \mathcal{I}_{B}) (\rho_L \rho_L^{\dagger} )(Q_A^{-1} \otimes \mathcal{I}_{B}).
    \nonumber
    \end{aligned}
\end{equation}
{\color{black}To deal with real parameters in NN models, the lower-triangular matrices can be further split into real and imaginary parts, ending up with a real vector $\boldsymbol{\alpha}$~\cite{ma2023attention,ma2023tomography,ahmed2021quantum,danaci2021machine}. 
As demonstrated in Fig.~\ref{fig:framework}, given the frequency vector calculated from the measurement outcomes of different measurement operators, the NN model is required to return a real vector $\boldsymbol{\hat{\alpha}}$ that corresponds to a physical density matrix via Eq.~(\ref{eq:tau2rho}). The ground truth of the vector $\boldsymbol{\alpha}$ is obtained via Cholesky decomposition of the density matrix $\rho$, which has a one-to-one correspondence to  $\rho$~\cite{ma2023attention,ma2023tomography,ahmed2021quantum,danaci2021machine}, 
Then, the parameters $\xi$ are trained to minimize the difference between the expected value $\boldsymbol{\alpha}$ and the predicted value $\boldsymbol{\hat{\alpha}}$. }
	
NNs are being widely used to characterize quantum states with various architectural designs. Fully connected neural networks (FCNs) were firstly adopted for QST~\cite{xu2018neural} and exhibited potential in denoising the SPAM errors~\cite{palmieri2020experimental} and the sampling noise due to limited measurement resources~\cite{Ma2021CDC,ma2024neural}. By converting measurement outcomes into images~\cite{wang2022ultrafast,lohani2020machine,danaci2021machine,lohani2021experimental}, CNNs have been effective in addressing challenges related to incomplete measurements and adaptive dimensions~\cite{lohani2023dimension}. Recent attempts have been to explore sequential information among quantum data~\cite{carrasquilla2019reconstructing,zhu2022flexible}, focusing on the similarities between quantum patterns and language structure. As demonstrated in Fig.~\ref{fig:language model}, quantum correlations exist at the level of one-shot measurements and the level of expectations of many measurements. Such a hierarchical structure resembles describing an entity using characters, words, and finally a sentence~\cite{ma2023attention}. {\color{black}In particular, the attention mechanism can be drawn to characterize long-range quantum entanglement among qubits (reflected in projective measurements of the quantum state)}, benefiting the task of learning the probability function of GHZ states with an order-of-magnitude improvement in the sample complexity compared to RNN-based tomography~\cite{cha2021attention,zhong2022quantum}. Another work notes the similarity between words and frequencies of a set of measurements and thus proposes a solution for QST: translating experimentally observed frequencies into physical density matrices, thus realizing a full tomography~\cite{ma2023attention}, exhibiting an order-of-magnitude improvement in the log of infidelity over FCN methods and CNN methods.

GANs offer a novel approach to learning the mapping between a latent space and data and have been extensively investigated for QST~\cite{braccia2022quantum,ahmed2021quantum,carrasquilla2019reconstructing,zhu2022flexible,ahmed2021classification}. In this context, QST is conceptualized as a generative adversarial game involving two players. The generator aims to produce data closely resembling the true data distribution, and the discriminator is trained to distinguish real data from fake data originating from the generator. For QST tasks, it is essential to introduce a variable to control the output, which is known as a conditional GAN. Let the conditional input vector be the composed elements of measured results and measurement operators $\boldsymbol{x}=[\mathcal{M},\boldsymbol{f}]$, and the noise be $\boldsymbol{z}$. The generator functions as a mapping $\{\boldsymbol{x},\boldsymbol{z}\} \rightarrow \rho$. A quantum version of generative adversarial learning has been theoretically proposed to exhibit an exponential advantage over its classical counterpart~\cite{lloyd2018quantum,braccia2021enhance}. Quantum GANs based on quantum superconducting circuits have been experimentally implemented to learn the properties of quantum states~\cite{hu2019quantum} with a fidelity of 98.8\% on average.

\subsection{ML-based process estimation}\label{SubSec:QPT}

Reconstructing the dynamics of quantum systems is important for establishing, for instance, channel fidelity in quantum communication, gate fidelity in quantum computing, and optimal parameter encoding for sensing applications~\cite{gebhart2023learning}. Without imposing any restrictions on quantum dynamics, we first introduce how ML can benefit quantum process tomography. Then we focus on Hamiltonian learning as an illustration. Finally, the dynamics of an open quantum system are investigated. 

	
\subsubsection{Quantum process tomography}
	
	
Recall that a quantum process can be defined as a completely positive map $\Lambda$ that transforms an input state $\rho^{\text{in}}$ to an output state $\Lambda(\rho^{\text{in}})$~\cite{nielsen2010quantum}. The task of fully reconstructing the underlying unknown dynamics of a quantum system is called quantum process tomography (QPT). In the standard approach, one estimates the dynamical process by applying it to a set of known quantum states, referred to as probe states $\{\rho_j^{\text{in}}\}$. The output state for each probe state, i.e., $\Lambda(\rho_j^{\text{in}})$ is then reconstructed via QST~\cite{zeek1999pulse}. To achieve a complete reconstruction of $\Lambda$, the probe states must span a basis for all possible initial states, and the measurements for QST should be tomographically complete. Consequently, full QPT presents greater challenges than QST~\cite{mohseni2008quantum}. 

 According to the Choi-Jamiolkowski isomorphism~\cite{choi1975completely}, there exists a one-to-one correspondence between every quantum map $\Lambda$ and a Choi operator $Q_{\text{Choi}}$. A normalized $Q_{\mathrm{Choi}}$ plays a similar role as a density matrix. This intrinsic analogy between QST and QPT enables all of the theorems about quantum maps to be derived directly from those of quantum states. For example, given $N_P$ probe states and $N_M$ measurements $\{\mathcal{M}_k\}$ for QST, we have the measured frequencies $f_{jk}$ with $j\in\{1,2,...N_P\}$ and $k\in\{1,2,...,N_M\}$. Treating $(\rho_{j}^{\text{in}})^T \otimes \mathcal{M}_k$ as an entity allows for the conceptualization of a quantum process as a quantum state within a larger Hilbert space. {\color{black}Let the dimension of the system be $d$, then the corresponding dimension of $(\rho_{j}^{\text{in}})^T \otimes \mathcal{M}_k$ is $d^2$. In principle, QPT can be naturally reduced to QST, for a small number of qubits}. Based on these observations,  QPT  can be simplified as approximating a function that maps $\{f_{jk}\}$ into $Q_{\textup{Choi}}$. Theoretically, various state estimation techniques can be applied to the characterization of quantum processes as well, including the NN architecture design in Subsection~\ref{subsec:ffn}.

{\color{black}When prior knowledge about the process $\Lambda$ is available, e.g., a unitary process with a fixed number of unknown parameters in system Hamiltonian, the number of free parameters in $\Lambda$ does not scale as $(2^{4n}-2^{2n})$. In such cases, the estimation can be improved beyond the limitation of exponential scaling of measurement resources, e.g., the number of measurement copies $N_M$. For example, there are attempts to reconstruct a unitary quantum process by inverting the dynamics using a variational algorithm~\cite{carolan2020variational,xue2022variational}. Under this framework, one can variationally train a quantum circuit to unravel the operation of an unknown unitary on a known input state, essentially learning the inverse of the black-box quantum dynamics. RNNs have been applied to learn the non-equilibrium dynamics of a many-body quantum system from its nonlinear response under random driving~\cite{mohseni2022deep}.  QGAN-based approximations of a quantum map $\Lambda$ have also been proposed to characterize spatially or temporally correlated noise in quantum circuits~\cite{braccia2022quantum}.}

\subsubsection{Hamiltonian learning}

{\color{black}
In many applications, we may be interested in identifying a unitary process, and only the system Hamiltonian needs to be characterized.
For a $d$-dimensional system, the problem is then to learn the appropriate $d\times d$-matrix $\mathcal{H}$. Several methods using the Fourier transform or fitting on the temporal records of measurement of some observables have also been proposed to estimate
Hamiltonians of few qubits~\cite{di2009hamiltonian}. In some cases, prior knowledge about the structure of the system Hamiltonian is available. Hence, one can consider a parameterized form of the Hamiltonian governing the quantum dynamics as $\mathcal{H} = \sum_m \mu_m X_m$, where $\mu_m$ is a vector consisting of unknown parameters. In these cases, it is possible to character many-body Hamiltonians using polynomial parameters. This allows for the estimation of the system Hamiltonian even if only a small fraction of the subsystems in the network, e.g., one or two can be measured. For example, protocols able to find the coefficients characterizing the interaction Hamiltonian have been developed, including the eigenstate realization algorithm~\cite{zhang2014quantum}. The capability of ML to capture patterns from measured data makes it suitable for solving quantum Hamiltonian learning that involves time-correlated data. }

\begin{figure}
		\centering
		\includegraphics[width=0.48\textwidth]{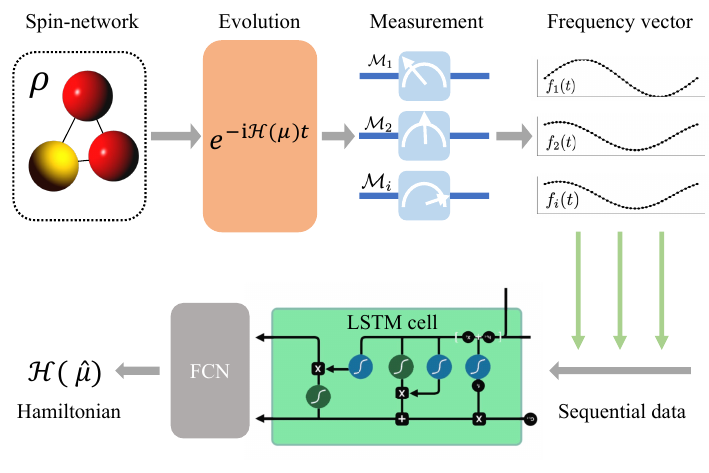}
		\caption{Diagram of NNs for learning the parameters of time-independent Hamiltonians from the temporal records of single-qubit measurements. {\color{black}Starting from the initial state $\rho$, the dynamical evolution $e^{-{\rm{i}}\mathcal{H}(\mu)t}$ is performed, during which the expectation values of single-qubit measurement operators $\{\mathcal{M}_i\}$ (e.g., single-qubit Pauli operators $\sigma_k^{(j)}$ with $k\in\{x,y,z\}$ and $j\in {1,2,...,3}$) are measured at each time step. These expectation values over a period are then collected into a frequency vector $f_i(t)=[f_i^{t_0},f_i^{t_1},...]$, which is fed into an LSTM cell to predict the parameters $\hat{\mu}$ of the Hamiltonian}.}
		\label{fig:QHL}
	\end{figure}

{\color{black} The prior of the system Hamiltonian allows for the identification of Hamiltonians under the temporal records of single-qubit measurements, e.g., system Hamiltonian can be characterized with polynomial parameters. In such cases, there exists the underlying rule from single-qubit measurements to the target Hamiltonians~\cite{che2021learning}. This can be learned via data-driven machine learning although this rule may have complicated or even unknown functional forms~\cite{che2021learning}. Motivated by this, LSTM (one variant of RNNs) has been introduced to learn the target Hamiltonians from the temporal records of single-qubit measurements~\cite{che2021learning}. As illustrated in Fig.~\ref{fig:QHL}, sequential data is injected into an LSTM block, and followed by an FCN to reconstruct a time-independent parameter $\hat{\mu}$. For time-independent Hamiltonian learning, one should replace the FCN (grey block) with an additional LSTM module to reconstruct sequential data, i.e., time-dependent parameters $\hat{\mu}(t)$. Strong robustness against measurement noise and decoherence effects has been observed when learning the magnitude and sign of parameters in Hamiltonians, for systems with up to 7 qubits.}

To go beyond the limitations of prior knowledge about the coupling structure of the original Hamiltonian~\cite{che2021learning}, physics-enhanced Heisenberg NNs are defined together with a physics-motivated loss function based on the Heisenberg equation, which ``forces” the NNs to follow the quantum evolution of the spin variables~\cite{han2021tomography}. In the extreme case in which measurements are taken from only one spin, the achieved tomography fidelity values can reach about 90\%.

\subsubsection{Learning open quantum system dynamics}
  When a quantum system is sufficiently isolated from its \emph{environment}, the Hamiltonian description discussed above is a good approximation of the quantum dynamics. For a general quantum process, the system should be treated as an open quantum system~\cite{breuer2002theory}. For Markovian quantum dynamics, its state evolution can be described by the MME in Eq.~(\ref{eq:lindblad}). Learning such Markovian dynamics can be much more challenging than Hamiltonian learning: recasting equation into the equivalent form $\rho=\mathcal{L}(\rho)$ shows that the task now involves learning a $d^2 \times  d^2$-dimensional Liouvillian superoperator $\mathcal{L}$ for a $d$-dimensional system. {\color{black}By representing the mixed many-body quantum states with complex-valued RBMs~\cite{hartmann2019neural}, the dissipative dynamics of one-dimensional lattices can be well approximated, and the resulting stationary states can be effectively characterized.}
  
	
For non-Markovian quantum dynamics, the state at time $t+\Delta t$, i.e., $\rho_{t+\Delta t}$, depends not only on $\rho_{t}$ but also on the system’s history at earlier times. Denote $\leq t$ as a superoperator that not only depends on $t$ but also the entire history before time $t$. A compact representation for non-Markovian dynamics $\dot{\rho}(t)= \mathcal{L}_{\leq t}[\rho]$ can be obtained as follows
\begin{equation}
  \dot{\rho}(t)=-\mathrm{i}[\mathcal{H}+\mathcal{H}_{\leq t}^{LS},\rho]+\sum_{\mu} [L_{\leq t}^{\mu} \rho L_{\leq t}^{\mu\dagger}-\frac{1}{2}\{L_{\leq t}^{\mu} \rho L_{\leq t}^{\mu\dagger},\rho\}],
  \label{eq:non-Markovian}
\end{equation}
 where $L_{\leq t}^{\mu} $ are (recurrent) Lindblad operators describing the coupling channel with the environment and $\mathcal{H}^{LS}$ is a ``Lamb-shift" term, namely a correction to the Hamiltonian induced by the environment~\cite{banchi2018modelling}. This is reasonable because for small enough $\Delta t$, the time evolution of a quantum state can be simply given as  $\rho(t+\Delta t) \approx e^{\Delta t\mathcal{L}_{\leq t}[\rho(t)]}$, which is a completely positive trace-preserving quantum map~\cite{banchi2018modelling}. To capture the long correlations between different time series, RNNs were used to model the long-range memory for non-Markovian dynamics. {\color{black}Directly learning from data also enables us to effectively model complex quantum correlations between systems and environments with a constant and fixed number of parameters. For example, RNN can be utilized to predict a single recurrent Lindblad $L_{\leq t}^{\mu}$ ($\mu=1$) and $\mathcal{H}_{\leq t}^{LS}$ (in Eq.~(\ref{eq:non-Markovian})) for a two-level system with spontaneous decay~\cite{banchi2018modelling}}. 
 CNNs have been utilized to predict non-Markovian reduced system dynamics in a broad range of dynamical regimes from weakly damped coherent motion to incoherent decay~\cite{herrera2021convolutional}. {\color{black}This approach yields small deviations (3.6\%) between the predicted and exact populations in 2-level quantum systems, while also reducing the computational resources required for long-time simulations.}

\subsection{Outlook and future directions}\label{Subsec:Summary}


ML methods have the potential to achieve improved estimation accuracy in many practical situations such as few copies and noisy measurement data~\cite{ma2023attention,lohani2020machine,danaci2021machine}. Several advantages of employing ML methods for quantum estimation include:
 1) we may not necessarily need complete measurement bases; 2) artificially generated data can be used to train the learner offline for quantum estimation tasks; 3) ML methods can be online integrated into developing adaptive quantum estimation strategies for enhancing estimation accuracy. Still, various challenges deserve further investigation which opens up new opportunities for future research.

\textbf{Model complexity and scalability}: {\color{black}Additional effort is needed to consider the involved parameters in NN models versus the number of parameters in quantum systems, e.g., ($4^n-1$) free parameters in an $n$-qubit density matrix. Existing achievements in full tomography mainly focus on low-qubit states. As quantum systems grow in complexity, scaling ML algorithms to efficiently process and analyze the increasing amount of data becomes more challenging.} Constructing an approximate classical description of a quantum state using very few measurements has been proposed as a classical shadow of quantum states~\cite{huang2020predicting}. It would be useful to investigate how to incorporate ML methods to efficiently capture shadows of quantum entities (e.g., quantum channels), therefore predicting the properties of large-scale quantum systems.

\textbf{Benchchmarks and accuracy}: {\color{black}Despite the capabilities of ML-based methods in different estimation tasks, it is usually difficult to characterize their accuracy (e.g., fidelity or mean squared error) versus model complexity (parameters in NN models and the resources used), which remains an interesting question. Although there are some numerical results to determine the scaling of accuracy versus measurement copies of ML-based estimation methods, e.g., Transformer-based QST in \cite{cha2021attention}, it would be useful to obtain an analytic solution to the scaling and whether the scaling could reach the fundamental limit $\sim \frac{1}{\sqrt{N_M}}$. Considering the additional training overhead in ML-based methods typically absent in traditional methods, it would be interesting to fairly compare the ML-based methods and classical methods with the full consideration of computational complexity.}

\textbf{Generalization}: Although ML-based estimation methods demonstrate robustness against different errors, their generalization performance across different types of quantum samples remains inferior to re-training for a new class of samples. This suggests a potential avenue for leveraging relationships between different tasks thus improving generalization. One promising approach is to employ advanced ML techniques such as transfer learning to reuse knowledge gained from previous quantum tomography tasks to improve performance on new, but related, tasks. Useful examples might include: 1) quantum tomography tasks with varying measurement settings, exploring the relationship between different measurement bases; 2) Transitioning knowledge gained from state estimation to closely related tasks such as process estimation by leveraging the similarity between density matrices and Choi matrices, or detector estimation by understanding the relative relationship between state and measurement.

\textbf{Real-time adaptive estimation}: Current algorithms for quantum estimation are costly both experimentally, requiring measurement of many copies of the state, and computationally, needing significant time to analyze the gathered data. Considering that the estimation of quantum entities relies on the selection of quantum measurement bases, it is desirable to adopt an adaptive way to adjust the measurements therefore achieving the estimation tasks with reduced samples. Hence, it would be interesting to leverage the power of ML to adapt measurement bases in real-time thus providing a fast and flexible approach for QT.

\section{Learning-based optimization for quantum control}\label{Sec:learningcontrol}
	
Quantum control aims to direct the evolution of quantum systems, with the objective often being to maximize a specific performance function~\cite{rabitz2000whither}. It can be often formulated as an optimization problem. Learning-based control is an effective approach that can learn from the previous experience and optimize the system performance by searching for the best control strategy in an iterative way. In the following, we first outline the process of converting quantum control into an optimization problem in Subsection~\ref{subsec:s3s2}, highlighting the role of gradient-based methods in addressing this challenge. Following this, we explore the application of evolutionary computing techniques for optimizing quantum systems in Subsection~\ref{subsec:s3s3}. We also discuss the experimental applications of learning-based optimization for quantum control in Subsection~\ref{subsec:s3s4}. The section concludes with a discussion of the challenges and prospects of learning-based optimization strategies in the realm of quantum control.

\subsection{Quantum control as an optimization problem}\label{subsec:s3s2}
	
The objective of quantum control problems can be usually formulated as an optimal control problem. This involves transforming the challenge into the task of optimizing a function, which depends on variables or control parameters such as quantum states, control inputs, and control time~\cite{dong2010quantum}. To systematically study the relationship between the time-dependent controls and the associated values of the objective functional, a notion of quantum control landscape~\cite{rabitz2004quantum,Chakrabarti-and-Rabitz-2007} is defined as the map between the time-dependent control Hamiltonian and associated values of the control performance functional. Given a control field, $u=\{u_m(t)\}$,  one may define a performance functional  $\Phi$, which may be a given functional of {\color{black}the state flow $|\psi(t)\rangle$} and the control defined according to the practical requirements~\cite{chen2013closed}. For example, the fidelity $\Phi=|\langle\psi(T)|\psi_f\rangle|^{2}$ between the final state $|\psi(T)\rangle$ and the target state $|\psi_f\rangle$ or the expectation $\Phi=|\langle\psi(T)| \mathcal{P} |\psi(T)\rangle|^{2}$ of an operator $\mathcal{P}$ may be defined as a performance index for a state transfer task. {\color{black}Define $\langle \mathcal{U}_f|\mathcal{U}(T)\rangle = \frac{1}{d}\textup{Tr}( \mathcal{U}_f^{\dagger} \mathcal{U}(T))$. For 
the optimal control problem of unitary transformations (e.g., quantum gates),  the performance function may be defined as $\Phi=|{\langle \mathcal{U}_f|\mathcal{U}(T)\rangle}|^2$~\cite{dong2016learning}}.

These problems can be solved using a unified framework of gradient-based methods, where the control fields are iteratively updated in the direction of the gradient of $\frac{\delta \Phi}{\delta u_{m}(t)}$ with a learning rate $\zeta$. Specifically, for a maximization problem, the control fields can be updated as follows:
	\begin{equation}
		u^{k+1}_{m}(t)=u^{k}_{m}(t)+\zeta \frac{\delta \Phi}{\delta u_{m}(t)}.
	\end{equation}
Following this idea, gradient ascent pulse engineering (GRAPE) was developed to maximize the performance $\Phi$ for various quantum control tasks~\cite{khaneja2005optimal}. {\color{black}Another popular method is called the Krotov method~\cite{jager2014optimal}, where combined information from forward and backward propagation is utilized to update the control fields. This method guarantees monotonic convergence and is well-suited for complex and constrained quantum control problems}. The open GRAPE algorithm has also been developed to calculate the gradient based on the master equation \cite{schulte2011optimal}. A gradient-based frequency-domain optimization algorithm has been developed to solve the optimal control problem with constraints in the frequency domain~\cite{Shu-PRA2016}.

In practical applications, robustness is an important requirement due to the existence of uncertainties. For example, an inhomogeneous quantum ensemble usually consists of numerous individual quantum systems (e.g., atoms, molecules, or spin systems), each characterized by parameters that may exhibit variations~\cite{Li-and-Khaneja-2006, Chen2014PRA}. These variations could manifest themselves as dispersion in the strength of the applied radio frequency field or fluctuations in the natural frequencies of spins in NMR systems~\cite{Li-and-Khaneja-2006}. To employ the same control fields that steer individual systems with different dynamics from a given initial state to a target state, a sampling-based learning control (SLC) method has been developed~\cite{Chen2014PRA}. An augmented system consisting of $N_S$ representative samples over the distribution 
can be constructed, which can be optimized according to the following average performance function
\begin{equation}\label{eq:cost}
\bar{\Phi}(u)=\frac{1}{N_S}\sum_{j=1}^{N_S} \Phi_{\omega^j}(u),
\end{equation}
 where $\Phi_{\omega^j}(u)$ represents the objective function for a given sample $\omega^j$. Applying the SLC idea to GRAPE, a sample-based gradient algorithm (s-GRAPE) has been developed, wherein the control fields can be updated as
$$u^{k+1}_{m}(t)=u^{k}_{m}(t)+ \zeta \frac{\delta \bar{\Phi}}{\delta u_{m}(t)}.$$ 

This approach holds promise for inhomogeneous quantum ensembles and quantum robust control~\cite{dong2016learning,dong2020learning}. Drawing inspiration from deep learning, a batch-based gradient algorithm (b-GRAPE) has emerged to delve into the richness and diversity of samples, thereby significantly enhancing the control robustness while preserving high fidelity~\cite{wu2019learning}. An adversarial gradient-based learning algorithm (a-GRAPE) can acquire highly resilient controls by generating adversarial samples through the pursuit of Nash equilibria~\cite{ge2020robust}. A data-driven gradient optimization algorithm (d-GRAPE) has been proposed to correct deterministic gate errors by jointly learning from a design model and the experimental data from quantum tomography~\cite{Wu2018dGRAPE} and this method can be further upgraded as c-GRAPE algorithm by combining adaptive QST from the experimental data~\cite{ding2021collaborative}.
 According to quantum control landscape theory~\cite{Chakrabarti-and-Rabitz-2007}, gradient-based learning methods typically excel in solving optimal control problems when the system model is known. However, this assumption may not always hold in experimental setups. To mitigate this challenge, one might resort to an evolutionary computation-based approach to seek effective solutions.

\subsection{Evolutionary computation for quantum control}\label{subsec:s3s3}
	
For a quantum control problem, the gradient-based methods typically excel provided that (i) obtaining the gradient is straightforward and (ii) there are no local traps on the control landscape~\cite{Chakrabarti-and-Rabitz-2007}. Nevertheless, ensuring these conditions for complex quantum systems is often challenging. In such cases, leveraging stochastic search algorithms becomes a natural choice for finding effective controls. Here, we delve into evolutionary computation, extensively utilized across various engineering domains, spanning from molecular to astronomical scales~\cite{eiben2015evolutionary}. Evolutionary computation algorithms draw inspiration from the natural selection process~\cite{eiben2015evolutionary}, where the most adept individuals are chosen for reproduction, thereby generating offspring via different variations for the subsequent generation. {\color{black}To implement this concept, it is essential to analogize potential solutions as individuals within a population and to establish a measure of ``fitness" based on the quality of the solutions. Consequently, the overall process can be outlined as a loop (see the optimization loop in Fig.~\ref{fig:close-loop}) of evaluating the current generation of solutions, then creating new solutions through different variations, and selecting some to act as the basis for the next generation. In the context of genetic algorithms (GA) and differential evolution (DE), the variation phase mainly comprises two crucial operations: ``mutation" and ``crossover"~\cite{zeidler2001evolutionary}}. 

	
When addressing quantum control tasks, the ``fitness" function for each vector corresponds to the functional $\Phi(u)$ for each control solution $u$~\cite{brown2023optimal}. Early attempts usually adopted GA to optimize the ``fitness" function of quantum control problems~\cite{judson1992teaching,bardeen1997feedback}. GA has also found applications in searching for control pulses for state preparation and quantum gate operations in nuclear magnetic resonance systems~\cite{manu2012singlet} and manipulating the ionization pathway of a Rydberg electron~\cite{gregoric2017quantum}. Meanwhile, DE has gained increasing attention in quantum control scenarios. For example, a subspace-selective self-adaptive DE variant has been developed to achieve a high-fidelity single-shot Toffoli gate (i.e., controlled-controlled-NOT (CCNOT) gate) and single-shot three-qubit gates~\cite{zahedinejad2015high,zahedinejad2016designing}. Despite sharing a similar mechanism, DE has been found to outperform GA and particle swarm optimization for ``hard" quantum control problems~\cite{zahedinejad2014evolutionary}, such as those requiring short durations for unitary operations or featuring a limited control parameter (for example, low $N_c$ in Eq.~(\ref{eq:Hamiltonian})). {\color{black}An improved DE algorithm introduces an efficient mutation rule that leverages information from both current and previous individuals. This approach has been validated on quantum state and gate preparation problems on 2-qubit NMR systems~\cite{yang2019improved}.} 

Unlike GA methods, which employ binary representation of candidate solutions and a low mutation probability~\cite{hegerty2009comparative}, 
DE methods represent solutions with real numbers and operates with a higher mutation rate~\cite{panduro2009comparison,zahedinejad2014evolutionary}. This enables DE to explore the search space more effectively, diminishing the risk of becoming trapped in local minima, which is particularly crucial for quantum control tasks~\cite{ma2015differential}. Another notable aspect of DE is its versatility in mutation strategy selection since several DE variants based on mixed strategies have exhibited good performance for different optimization tasks~\cite{qin2008differential,mallipeddi2011differential,becerra2006cultured}. When it comes to the context of quantum control problems, DE with a single strategy may suffice for simple quantum control problems, while DE variants with mixed strategies may be a promising candidate for quantum control problems with multimodal landscapes~\cite{ma2017CTT}. To facilitate this, one can construct a strategy pool consisting of several mutation schemes with effective yet diverse characteristics. For example, it has been found that four strategies can yield favorable performance for controlling open quantum systems~\cite{ma2015differential} with high fidelity with uncertain parameters considered, as well as achieving consensus in quantum networks (all nodes in a network hold the same substates~\cite{dong2019learning}).
	
In applications where the robustness of the control fields is required, one may either use Hessian matrix information~\cite{xing2014assessment} or integrate the concept of SLC into the learning algorithm~\cite{palittapongarnpim2017learning,dong2019learning}. Compared with gradient-based methods, DE performs much better when imperfections and measurement errors are involved~\cite{yang2020assessing}. Improvements in DE, such as dynamic parameter variation for mutation and crossover~\cite{ma2017CTT,dong2019learning} and the introduction of a direction-adaptive mutation strategy, have resulted in improved robustness and faster processing in handling uncertainties like pulse imperfections and measurement errors~\cite{yang2019improved}. Furthermore, formulating quantum robust control as a multi-objective optimization problem has led to a two-step optimization strategy that prioritizes average fidelity before addressing infidelity variance, thereby bolstering solution robustness~\cite{hu2023two}.


\subsection{Adaptive learning control for quantum experiments}\label{subsec:s3s4}
	
	\begin{figure}
		\centering
		\includegraphics[width=0.48\textwidth]{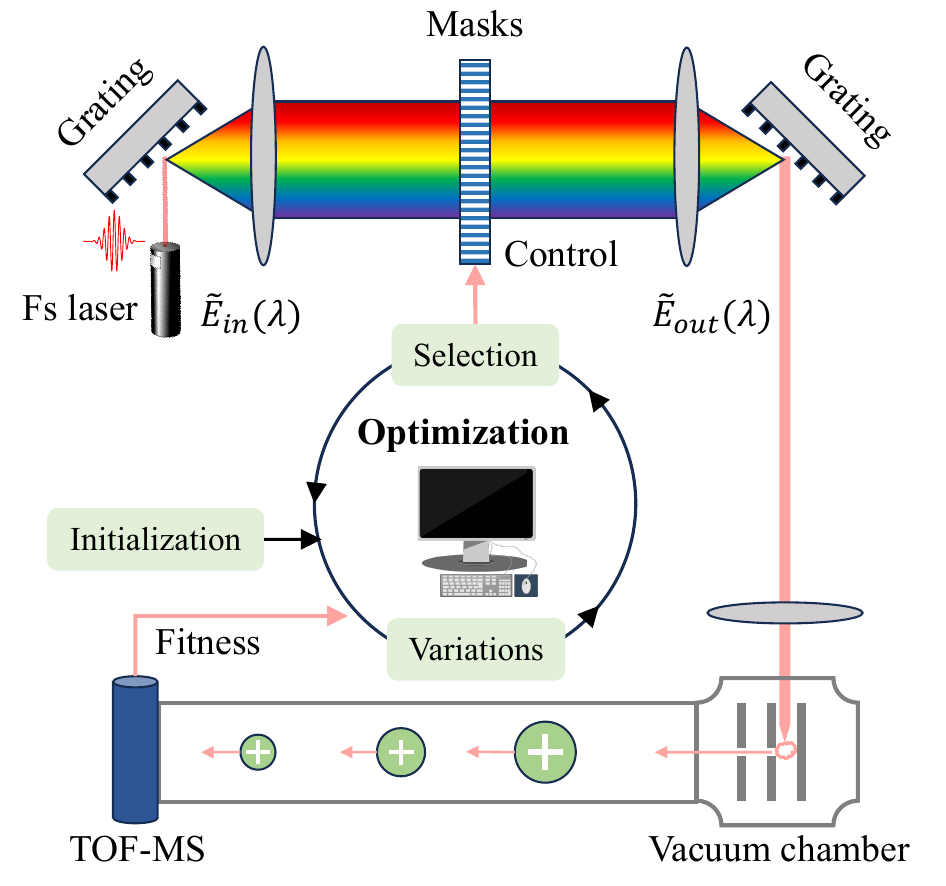}
		\caption{Illustration of the experimental setup of the femtosecond (fs) laser system using adaptive learning algorithms. {\color{black}The laser pulses are introduced into a pulse shaper that is equipped with a programmable dual mask liquid crystal spatial light modulator, where pulse shaping is typically achieved by modulating the phase and/or amplitude (i.e., the control fields to be optimized in the inner loop) of the laser frequency components with a computer-programmable spatial light modulator. The shaped laser pulses out of the shaper are focused into the vacuum chamber, where molecules will undergo ionization and dissociation, and their charged products can be separated and detected with a time-of-flight mass spectrometry (TOF-MS). In the inner optimization loop, the variations aim to perform genetic perturbations in the individuals, e.g., ``mutation" and ``crossover" in GA and DE~\cite{zeidler2001evolutionary}.}}
		\label{fig:close-loop}
	\end{figure}

When implementing learning methods in experimental quantum systems, the control fields undergo iterative updates to maximize control performance~\cite{brif2010control}. Since its introduction, GRAPE  has demonstrated wide applications in NMR systems, particularly in modules for state preparation~\cite{filgueiras2012experimental,lu2010experimental}. These applications often necessitate computing numerous time propagations of the controlled system's state, presenting challenges for classical computers, especially in handling high-dimensional systems. To overcome this limitation, researchers have developed methods to approximate the ``fitness" function and its gradient for control inputs through evolutionary and measurement processes on a quantum simulator. This approach has facilitated the experimental preparation of complex quantum states, such as a 7-correlated quantum state~\cite{li2017hybrid} and a 12-coherent state~\cite{li2017hybrid}. Experimental verification has been conducted on a solid-state ensemble of coupled electron-nuclear spins~\cite{feng2018gradient}. Recently, an iterative GRAPE algorithm has been proposed to decompose large-scale problems into a set of lower-dimensional optimization subproblems through disentanglement operations, with experimental verifications on a 4-qubit NMR system~\cite{chen2023accelerating}.

Another groundbreaking advancement is the selective laser modulation of physical and chemical phenomena, enabling the production of numerous samples in identical states for laboratory chemical molecules~\cite{rabitz2000whither}. Those experiments generally involve three elements: 1) designing a trial control input, 2) generating and applying this control to a sample in a laboratory setting to observe its effects, and 3) employing a learning algorithm that leverages data from previous experiments to update parameter setting to generate the new control pulses. This can be solved using a closed-loop learning control approach~\cite{chen2013closed}.  A notable feature of this approach is its resilience to initial trials, enabling the generation of laser pulses simply and effectively, without necessitating prior knowledge of the input pulse shape~\cite{chen2013closed}. Given an appropriate control objective, the key lies in adopting an effective learning algorithm, which is sufficiently efficient for searching for an optimal control pulse~\cite{rabitz2000whither}.
	
	
As demonstrated in Fig.~\ref{fig:close-loop}, a setup for the femtosecond (fs) pulse-shaping experiment usually consists of a laser, a set of molecules, and a measurement device~\cite{rabitz2000whither}. The apparatus acts as an input-output device capable of reliably reporting the  \emph{action} of any introduced field upon the molecules and is coupled to a learning algorithm capable of recognizing patterns in the input-output measurement relationships and thus guiding an iterative sequence of new experiments. The iteration is facilitated by a cost function that only contains costs for the target state and laboratory considerations (e.g., constraints on the form of the field). Following this framework, teaching a laser pulse sequence to excite specified molecular states has been experimentally realized using a GA method to search for good solutions in parameter space~\cite{judson1992teaching}. Other achievements include the maximal compression of fs laser pulses~\cite{baumert1997femtosecond,yelin1997adaptive,brixner1999feedback,zeek1999pulse,zeek2000adaptive}, e.g, shortening the time width of laser pulses and increasing their peak power for various applications. The adaptive mechanism makes it possible to generate maximally compressed laser pulses simply and effectively, without requiring knowledge of the input pulse’s shape. Meanwhile, DE has also been employed for selective control of molecular fragmentation~\cite{dong2019learning}. In this experiment, $\text{CH}_{2}\text{BrI}$ molecules undergo ionization and dissociation, and their charged products can be separated and detected with a time-of-flight mass spectrometry (TOF-MS). The photoproduct ratio of $\text{CH}_2\text{Br}^{+}/\text{CH}_2\text{I}^{+}$ was chosen as the control objective, which corresponds to breaking the weak C-I bond versus the strong C-Br bond~\cite{dong2019learning}. This method has also been applied to fragmentation control of $\text{Pr(hfac)}_3$ using a fs laser~\cite{dong2020learning}.

The adaptive control of quantum experiments has succeeded in compressing broadband laser pulses on semiconductors~\cite{kunde2000adaptive,kunde2001optimization,siegner2002adaptive}. The use of phase-modulated fs laser pulses to exploit semiconductor nonlinearities has been explored to develop an ultrafast all-optical switch~\cite{kunde2000adaptive,kunde2001optimization}. Through evolutionary strategies, researchers have determined optimal pulse shapes that significantly enhance the ultrafast semiconductor nonlinearities, nearly quadrupling their effect. This technique has been further applied to coherently control two-photon-induced photocurrents in two distinct types of semiconductor diodes~\cite{chung2006coherent}. Moreover, the field has seen advances in the optimal amplification of chirped fs laser pulses~\cite{efimov1998adaptive,efimov2000minimization}, demonstrating the broad applicability and effectiveness of adaptive control in quantum experiments.

	
	\subsection{Outlook and future directions}

Learning-based optimization of quantum control is a cutting-edge area that focuses on learning techniques to optimize the control of quantum systems. Several advantages of employing a learning-based approach for quantum control include: 1) it can be effective without knowledge of the quantum system dynamics; 2) it allows for easy implementations in experimental settings; and 3) the adaptive learning approaches bring robustness against possible uncertainties in quantum systems. However, it involves several challenges and opens up various future directions for research and development.
	
	
\textbf{Data efficiency}: For each learning trial, a fresh set of quantum ensembles is prepared to obtain the ``fitness" function which can be costly in experimental settings. Population-based methods involve evaluating numerous data points to suggest better solutions, often discarding past individuals and only retaining the current individuals and their associated ``fitness". It could be advantageous to store past information in a smart memory, providing insights for future individuals without recalculating ``fitness" from scratch. This approach would significantly reduce computational resources and is particularly beneficial for costly experimental implementations.
	
\textbf{Generalization}: Although control fields discovered through learning-based approaches exhibit robustness against errors in quantum control problems, they are typically tailored for a specific quantum system or task (e.g., a given initial or target state, or a fixed time duration for control pulses). For different problems, the common practice is to start learning from scratch, as the performance of directly applying existing control strategies often degrades. It is highly desirable to consider the similarities between different problems and design control strategies that generalize well across various quantum tasks. This issue, while challenging, is essential for achieving widespread applicability.

\textbf{Real-time implementation}. When applying this approach to experimental devices, additional efforts are required to integrate the learning routine into the entire system, such as using LabVIEW software. This integration can sometimes limit the overall efficiency. Given the popularity and versatility of these methods, it would be useful to design dedicated hardware, such as field-programmable gate array (FPGA) chips, to incorporate these algorithms directly into the devices and improve their sampling capability. This would enhance efficiency and streamline the process.

\section{Reinforcement learning for quantum control}\label{Sec:feedback}

  RL methods offer a considerable advantage in controlling systems without prior knowledge about the \emph{environment} and can be naturally applied to quantum control problems~\cite{sutton2018reinforcement}. In particular, RL techniques offer several advantages for quantum control tasks~\cite{giannelli2022tutorial}. They can handle complex and high-dimensional quantum systems~\cite{mnih2015human}, optimize control policies in real-time~\cite{sivak2022model}, and adapt to unknown or changing \emph{environment}s~\cite{taylor2009transfer}. In the following, we first briefly explain how to transform quantum control problems into a decision-making process in Subsection~\ref{Subsec:game}. Then, we investigate the utilization of RL methods in state-aware quantum tasks in Subsection~\ref{Subsec:aware}. After that, we turn to the case of partial observation in Subsection~\ref{Subsec:QEC}, followed by the investigation of quantum error correction in Subsection~\ref{Subsec:QEC}. Finally, we outline future directions for RL for quantum control in Subsection~\ref{Subsec5:summary}.

\subsection{Quantum control as a decision-making process}\label{Subsec:game}


In the traditional approaches to quantum control, the underlying model of quantum systems is often described using the system’s Hamiltonian or Schrödinger equation~\cite{dong2010quantum}. This allows for gradient-based optimization of the cost function~\cite{khaneja2005optimal}. In contrast, model-free approaches do not explicitly model quantum systems but instead rely on feedback signals from the experimental apparatus~\cite{sutton2018reinforcement}. They achieve optimization at a higher level by trial-and-error learning of \emph{action}-\emph{reward} patterns. RL approaches offer the distinct advantage of not requiring prior knowledge of complex systems, which has led to extensive investigation and applications in various quantum tasks.

\begin{figure}
\centering
\includegraphics[width=0.45\textwidth]{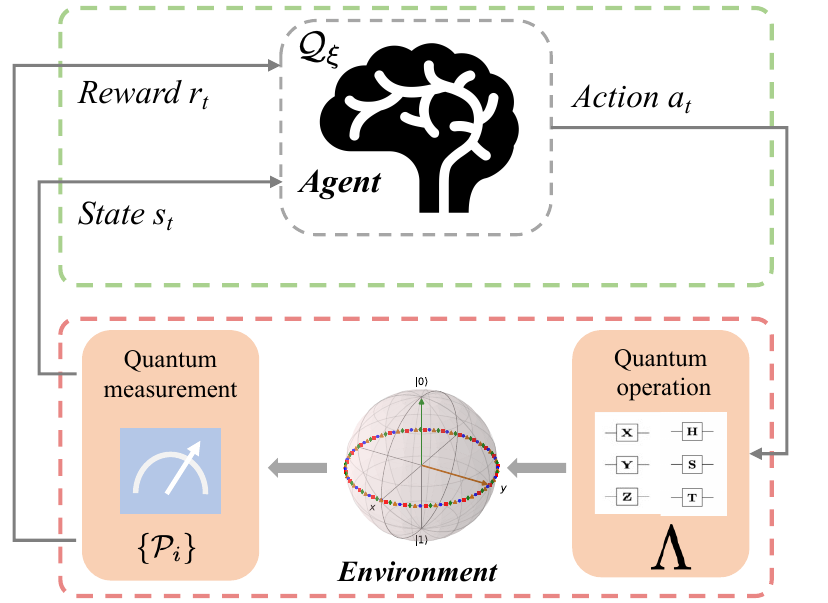}
\caption{Interaction between an \emph{agent} and \emph{environment} for one step of an RL task for quantum control. {\color{black}The \emph{agent} in the dashed green (classical) part can be represented as suggesting actions based on current states, i.e., $a_t=\mathcal{Q}_{\xi}(s_t)$. The \emph{environment} in the dashed orange (quantum) part is usually a quantum system that is subject to quantum operations for performing \emph{actions}, and quantum measurements for obtaining \emph{reward} signals. Here a Bloch sphere representation of a qubit is used as an example for illustration.} }
\label{fig:RLQC-framework}
\end{figure}

 {\color{black}The process of finding a control policy can be summarized as an \emph{agent} (dashed green part in Fig.~\ref{fig:RLQC-framework}), aiming to suggest a good action based on the current state, i.e., $a_t=\mathcal{Q}_{\xi}(s_t)$, with $\xi$ representing parameters to be optimized. In the context of different RL methods, the control policy of the \emph{agent} can be defined as $a_t=\pi_{\xi}(s_t)$ for policy-based methods or $a_t=\max_{a \in \mathbb{A}} Q_{\xi}(s_t,a)$ for value-based methods. The \emph{environment} refers to the quantum system to be investigated (dashed orange part in Fig.~\ref{fig:RLQC-framework}).} It is worth emphasizing the difference in the use of the term ``environment”: in quantum physics, it typically refers to a dissipative bath coupled to a quantum system, while in the RL context, it refers to the quantum system itself, which serves as the \emph{environment}. Quantum control problems can be formulated as a decision-making process. As demonstrated in  Fig.~\ref{fig:RLQC-framework}, upon observing the current \emph{state} $s_t\in \mathbb{S}$ (e.g., vector representations of the current quantum state $\rho(t)$, expectation values of some measurement operators, or one-shot measurement outcomes of the quantum system), an \emph{action} $a_t$ (e.g., a set of control $\{u_m(t)\}$ in Eq.~(\ref{eq:Hamiltonian}) or a choice from a set of quantum gates) recommended by the RL \emph{agent} is performed on the quantum systems, formulated as a quantum operation $\Lambda$. Based on the (usually unknown) dynamics of quantum systems, the next \emph{state} $s_{t+1}\in \mathbb{S}$ is obtained, along with a \emph{reward} $r_t \in \mathbb{R} $ obtained through quantum measurements. Finally, the tuple $(s_t,a_t,r_t,s_{t+1})$ consists of a transition.
 
 It is worth highlighting that the design of appropriate \emph{reward} functions plays a crucial role in guiding  RL \emph{agent} towards effective policies~\cite{ma2022curriculum}. In quantum control tasks, one can set \emph{reward} signals based on fidelity information $\mathcal{F}_t$, with non-linear transformation functions, e.g., $r_t=-\log_{10}(1-\mathcal{F}_t)$~\cite{chen2013fidelity,ma2022curriculum} or $r_t=\mathcal{F}_t^j$~\cite{porotti2022deep}, where $j$ could be a positive integer to control the non-linear transformation. {\color{black}By iteratively exploring the \emph{environment} with the past transitions collected, the parameters $\xi$ are updated to maximize the cumulative \emph{reward} among a sequence of transitions $\sum_{t =0} \gamma^t r_{t+1}$~\cite{sutton2018reinforcement}. As such, the \emph{agent} $\mathcal{Q}_{\xi}$ can discover optimal control strategies that lead to high performance (e.g., high fidelity) for quantum systems~\cite{niu2019universal}. In practical scenarios, the observations from quantum systems usually involve high-dimensional problems, creating a strong need for deep RL methods, such as trust region policy optimization (TRPO)~\cite{schulman2015trust} and proximal policy optimization (PPO)~\cite{schulman2017proximal} to tackle quantum tasks~\cite{porotti2019coherent,haug2020classifying,wauters2020reinforcement,sivak2023real,porotti2022deep,guo2021faster,an2021quantum,ding2021breaking,chen2019extreme}.}

  %


 In quantum contexts, the observation of a quantum system would be described by a measurement mapping in a state space model that depends on the current state and the previously applied action. Those situations fall into the paradigm of partially observable MDPs are proposed, where the observation is dependent on the current \emph{state} and the previous \emph{action}~\cite{kaelbling1998planning}. In particular, the expectation values of a projector $\mathcal{P}_i$, i.e., $p_i=\rm{Tr}(\rho \mathcal{P}_i)$ or even the one-shot measurement outcome of  $\mathcal{P}_i$, i.e., $\{0,1\}$ represent the partial \emph{state} of the \emph{agent}. Taking into account the specific characteristics of quantum systems, such as the challenge of obtaining a full description of quantum states, the applications of RL to quantum systems can be divided into two aspects: 1) \emph{state-aware} quantum tasks where full observability of quantum systems are available, in which $s_i$ can be obtained from $|\psi\rangle$ or $\rho$ (e.g., via splitting imaginary and real parts or other transformations). {\color{black}One should note that the acquisition of full knowledge about quantum states is reasonable when system models and initial states of quantum systems are known or informationally complete measurements on a large scale of identical copies are available for QST}; and 2) quantum tasks with \emph{partial observability}, where only measurement expectations or one-shot measurement outcomes of measurement operators are available. 



\subsection{RL for state-aware quantum tasks}\label{Subsec:aware}
	
RL techniques offer several advantages for quantum control tasks, including their ability to manage complex and high-dimensional quantum systems~\cite{an2021quantum}, optimize control policies in real-time~\cite{sivak2023real}, and adapt to unknown or dynamic \emph{environment}s (i.e., quantum systems)~\cite{giannelli2022tutorial}. In the early stages, Q-learning was utilized to identify variational protocols with nearly optimal fidelity~\cite{chen2013fidelity}, even in challenging situations, such as the glassy phase~\cite{bukov2018reinforcement} and quantum optics experiments~\cite{melnikov2018active}. In recent decades, various proposals have emerged for applying DRL to a wide range of quantum control problems. These include quantum state preparation~\cite{chen2013fidelity,zhang2019does,fosel2018reinforcement,zhang2018automatic,haug2020classifying,wu2020quantum,metz2023self}, quantum gate construction~\cite{an2019deep,an2021quantum,niu2019universal,daraeizadeh2020designing}, quantum metrology~\cite{xu2019generalizable,xu2021generalizable,qiu2022efficient,schuff2020improving,fallani2022learning}, quantum simulation~\cite{bolens2021reinforcement}, quantum spin squeezing~\cite{chen2019extreme,tan2021generation} and quantum approximate optimation algorithm (QAOA)~\cite{yao2022noise,jiang2022robust,wauters2020reinforcement}. Here we use two classes of quantum control problems to demonstrate the applications of RL: (i) coherent quantum control, and (ii) measurement-based feedback quantum control. Please refer to Table~\ref{table:DRLdifferentapplication} for the specific applications of RL methods.

\begin{figure}
		\centering
		\includegraphics[width=0.48\textwidth]{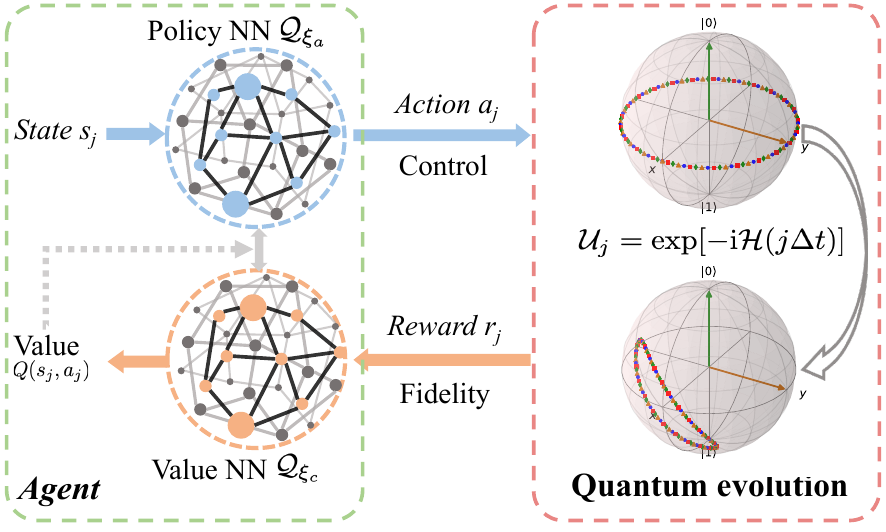}
		\caption{{\color{black}Schematic of Hamiltonian-based quantum control using actor-critic methods. At the iteration step $j$ with the current state $s_j$, the policy NN suggests an \emph{action} $a_j=\mathcal{Q}_{\xi_a}(s_j)$ that determines the current Hamiltonian of quantum systems to be $\mathcal{H}(j \Delta t)$. The equivalent unitary transformation $\mathcal{U}_j$ drives the quantum system into the next state $s_{j+1}$ with a \emph{reward} signal collected as $r_j$. The value NN evaluates the state-action values, i.e., $Q(s_j,a_j)=\mathcal{Q}_{\xi_c}(s_j,a_j)$, which provides implications for the optimization of RL \emph{agent}. In particular, ${\xi_c}$ is updated to minimize the mean squared error between the predicted value and the target value of $Q(s,a)$~\cite{mnih2015human}. Meanwhile, $\xi_{a}$ is updated to maximize the policy gradient, which relies on $Q(s,a)$~\cite{sutton2018reinforcement}.}}
		\label{fig:DRL-Hamitonian}
	\end{figure}

	\textbf{Coherent quantum control (Hamiltonian control).}
	Let us consider a quantum system with the Hamiltonian $\mathcal{H}(t)=\mathcal{H}_0+\sum_m u_m(t) \mathcal{H}_m$. Within this framework, the goal of RL is {\color{black}to discover a set of sequential control pulses $u(t)=\{u_m(t)\}_{m=1}^{N_c}$} to drive the quantum system to yield optimal performance. 
	Piece-wise control fields are widely used where the control is fixed during a short duration $\Delta t$. In this regard, for each time step $j$, we consider the system Hamiltonian to be constant in $[j\Delta t,(j+1)\Delta t]$. According to Eq.~(\ref{eq:Hamiltonian}), the coherent quantum control for time duration $\Delta t$ amounts to a unitary transformation $$\mathcal{U}_{j}=\exp[-{\rm{i}}\Delta t(\mathcal{H}_0+\sum_m u_m (j\Delta t)\mathcal{H}_m)].$$
	 As demonstrated in Fig.~\ref{fig:DRL-Hamitonian}, at each time step $j$ with the current \emph{state} $s_j$ (from $\rho(j\Delta t)$ or $|\psi(j\Delta t)\rangle$), the policy NN suggests an \emph{action} $a_j$ (corresponding to $u_n(t)$) that contributes to a system Hamiltonian $\mathcal{H}(j\Delta t )=\mathcal{H}_0+\sum_m u_m(j\Delta t)\mathcal{H}_m$. Then the quantum system evolves into the next state according to the unitary transformation $\mathcal{U}_j$ associated with the current Hamiltonian $\mathcal{H}( j \Delta t )$. Possible observation of quantum systems yields some \emph{reward} signals $r_j$, e.g., fidelity. Meanwhile, the transition $(s_j,a_j,s_{j+1},r_{j})$ is collected for updating the parameters in the RL \emph{agent}, and injected into both the policy NN and the value NN to update their parameters~\cite{sutton2018reinforcement}. 

	
This proposal has been extensively explored in different tasks. {\color{black}For example, DRL approaches have been utilized to learn all the driving protocols for global state preparation (over the continuous two-dimensional subspace
represented by the Bloch sphere, embedded in a higher-dimensional Hilbert space)~\cite{haug2020classifying}}. This approach
automatically finds clusters of similar protocols, which could be used to identify patterns and physical constraints in the protocols.
{\color{black}The Asynchronous Advantage Actor Critic (A3C) algorithm demonstrates its effectiveness for different atom number cases from $10$ to $10000$ without reforming the NNs and other parameters, contributing to an efficient and robust entanglement generation for quantum metrology within a short time duration}. The ultimate precision bounds exhibit the Heisenberg-limited scaling. 
A TRPO method has been employed to simultaneously optimize the speed and fidelity of quantum computation against both leakage and stochastic control errors~\cite{niu2019universal}. The potential of DRL has also been demonstrated in faster state preparation across a quantum phase transition \cite{guo2021faster} and robust digital quantum control with the operation time bounded by quantum speed limits dictated by shortcuts to  adiabaticity~\cite{ding2021breaking}. 
Enhanced DRL techniques have contributed to improving state preparation in different quantum systems, e.g., a faster transfer than that obtained with standard Gaussian pulses in an array of semiconductor quantum dots~\cite{porotti2019coherent} and the manipulation of Ag adatoms on Ag(111) surfaces with high precision, reaching success rates 95\% after more training under the new tip condition~\cite{chen2022precise}.


 When focusing on a sequence of quantum gates $\mathcal{U}_1, \mathcal{U}_2,..., \mathcal{U}_j,...$ rather than a sequence of control pulses, the problem of finding control pulses that achieves a desired transformation can be simplified. It involves decomposing one gate into a sequence of elementary gates, i.e., a finite universal set~\cite{moro2021quantum}. Given several elementary gates (e.g., H, S, T, CNOT,...), the \emph{agent} aims to select from the above candidate pool to realize a quantum gate capable of performing a desired task. For example,  an arbitrary single-qubit gate can be compiled into a sequence of elementary gates from a finite universal set~\cite{zhang2020topological,bolens2021reinforcement}. An RL-based quantum compiler has been developed to realize 2-qubit operators compiling~\cite{chen2024efficient}.

\textbf{Measurement-based feedback quantum control}. 
 In quantum feedback control, measurements on a quantum system generally perturb the system's state, introducing measurement-induced noisy dynamics, commonly known as quantum back  \emph{action}~\cite{nielsen2010quantum,sayrin2011real}. Recent efforts have been made to evaluate the performance of optimized feedback or adaptive measurement protocols using RL techniques. For example, RL has shown its capacity to effectively learn counterintuitive strategies for cooling a double well system to a state closely resembling a ``cat" state, exhibiting high fidelity with the true ground state~\cite{borah2021measurement}. Using a state-of-the-art DRL approach, measurement feedback control can be realized to produce and stabilize Fock states in a cavity subject to quantum-non-demolition detection of photon number~\cite{porotti2022deep}. Compared to traditional methods that rely on control Lyapunov functions for state stabilization, the DRL-based method works well without prior knowledge of quantum models. 
	
	\begin{table*}
		\centering 
		\caption{DRL for quantum control applications.}
		\begin{tabular}{ccc}
			\hline
			\textbf{Quantum applications}                                              &  \textbf{Control-level}       & \textbf{Adopted RL Methods}                       \\ \hline
			Quantum state preparation  &  Hamiltonian-control & 
			DQN\cite{mackeprang2019reinforcement,zhang2019does,ma2022curriculum}, Policy-gradient\cite{zhang2019does}, PPO\cite{haug2021machine}\\                                                             \hline   
			Quantum gate control & Hamiltonian-control &      DQN\cite{an2019deep}, Policy-gradient\cite{baum2021experimental} 
			\\ \hline
			
			Extreme spin squeezing & Hamiltonian-control & PPO\cite{chen2019extreme}
			\\ \hline
			Quantum metrology & Hamiltonian-control & A3C\cite{xu2019generalizable,qiu2022efficient}, DDPG\cite{xu2021generalizable}
			\\ \hline 
			Quantum compiler & Gate-control & DQN\cite{zhang2020topological,chen2024efficient}
			\\ \hline
			Quantum state engineering & Measurement-based & DQN\cite{mackeprang2019reinforcement}
			\\ \hline
			Quantum state stabilization & Measurement-based & DQN\cite{borah2021measurement}, PPO\cite{sivak2022model}
			\\ \hline 
			Quantum error correction & Measurement-based & Policy-gradient\cite{fosel2018reinforcement}, PPO\cite{sivak2022model,sivak2023real}
   \\ \hline 
		\end{tabular}
		\label{table:DRLdifferentapplication}
	\end{table*}

	
\subsection{RL for quantum control with partial observation}\label{Subsec:partial}
 Many existing results on RL for quantum control assume access to complete knowledge of quantum systems, which may be experimentally infeasible due to the exponential scaling of required quantum measurements on the number of qubits in many applications. Such prerequisites stand in contrast to the inherent properties of quantum stochasticity, and partial observability~\cite{bukov2018reinforcement,bilkis2020real,borah2021measurement,haug2021machine,sivak2022model}. Despite the limited knowledge available for quantum systems, RL \emph{agent}s do not depend on a deep understanding of the dynamics of quantum systems; {\color{black}instead, they focus on learning the patterns in the  \emph{action}-\emph{reward} relationship in a data-driven way. This highlights the fundamental challenge of applying RL to quantum physics: carrying out the optimization with stochastic data obtained from quantum systems.}

\begin{figure}
		\centering
		\includegraphics[width=0.48\textwidth]{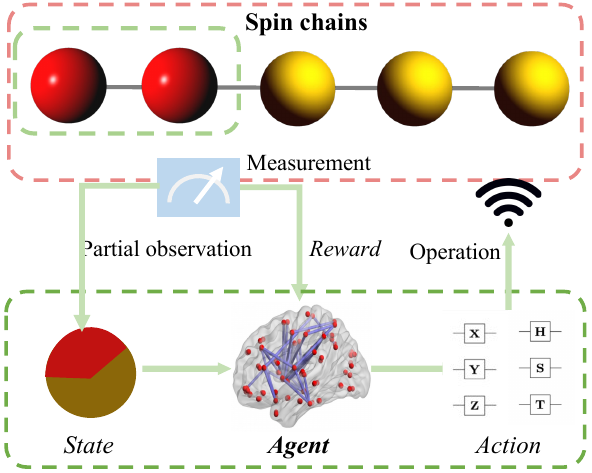}
		\caption{Schematic of RL for quantum control with partial observation. The \emph{agent} only has access to a partial observation of quantum systems. \emph{Reward} signals are obtained by performing measurements on quantum systems, which in turn influences the state of the quantum system. With complex dynamics hidden from the \emph{agent}, it can learn a state- \emph{action} pattern from the collected experiences.}
		\label{fig:DRL-partial-observation}
	\end{figure}

 A general framework for the application of RL to quantum systems with partial observation can be illustrated in Fig.~\ref{fig:DRL-partial-observation}, where quantum measurements (e.g., projective measurements) are utilized to capture limited knowledge about the system throughout the process. These intensive measurements raise several challenges: (i) Partial observations of quantum states, such as statistic-measured frequencies for corresponding measurements~\cite{baum2021experimental} or quantum properties (e.g., coherence, entanglement), can be collected and represented as the partial \emph{state} fed into the RL \emph{agent}. The reduced information in the partial \emph{state} representation might hinder the performance of learning by trial-an-error. (ii) The measurement process causes a random discontinuous jump in the underlying state~\cite{nielsen2010quantum}. 

Despite the inherent challenge of partial observability in quantum systems ~\cite{fosel2018reinforcement}, often referred to as the ``state-aware" issue~\cite{fosel2018reinforcement}, RL has proven itself to be a versatile tool to learn directly from stochastic measurement outcomes or low-sample estimators of physical observables.  For example, equipped with expectation values of the adjacent pairs of a 128-spin Ising chain, an RL \emph{agent} has successfully devised policies converging to the optimal adiabatic solution for QAOA~\cite{wauters2020reinforcement}. Given only incomplete Bloch vector representation (i.e., expectations of partial elements among the generalized complete Pauli operators, $\rm{Tr}(\rho \sigma_k), k \in\{x,y,z\}$), RL methods have attempted to realize high-fidelity quantum state preparation and QAOA applications~\cite{jiang2022robust}. Furthermore, an RL \emph{agent} has been designed to learn from the estimated density matrices based on measurement outcomes via QST, enabling the experimental realization of single-qubit gates on a superconducting quantum computer without prior knowledge of a specific Hamiltonian model, control parameters, or underlying error processes~\cite{baum2021experimental}. Researchers have taken a step further to explore the utilization of one-shot measurements rather than statistical values. {\color{black}For example, a \emph{state-aware} model using simulated data assists in the effective training of a \emph{state-unaware} model that suggests control based on experiment data of one-shot measurement outcomes~\cite{fosel2018reinforcement}.} Recently, efforts have brought RL closer to quantum observability, introducing an RL framework that relies exclusively on measurement outcomes as the sole source of information about the quantum state~\cite{sivak2022model}. An enhanced RL \emph{agent} has been experimentally trained using sole measurement data to initialize a qubit with real-time feedback from superconducting systems~\cite{reuer2023realizing}, where a low-latency NN architecture was specifically designed to process data concurrently with its acquisition, on a field-programmable gate array. In such settings, the \emph{state} representation may have a variable length, considering the temporal structure of the control sequences. Then, it is beneficial to utilize advanced architectures like LSTM~\cite{sivak2022model,sogabe2022model} and Transformer~\cite{vaswani2017attention}.


Apart from the low-sample state representations caused by partial observation, other critical aspects should be specifically considered when implementing the RL approach. {\color{black}Given the inherent stochasticity in quantum mechanics~\cite{nielsen2010quantum}, POMDPs rely on partial observations to suggest a good policy, which usually favors a stochastic policy rather than a deterministic policy~\cite{spaan2012partially}. This approach involves executing each experimental run with a different policy candidate, which can be then assessed using a binary \emph{reward}}. Unlike a deterministic policy, a stochastic policy capable of generating probabilistic \emph{action} may compensate for the randomness inherent in the quantum measurement process. For example, algorithms like PPO which involves comparing multiple policy candidates and performing small updates within the trust region, have found wide applications in various quantum tasks~\cite{chen2019extreme,an2021quantum,haug2020classifying,haug2021machine,guo2021faster,ding2021breaking,porotti2022deep,sivak2022model,sivak2023real}. 
	
Previous works usually define the \emph{reward} signal in terms of state fidelity, which requires a substantial number of quantum measurements to determine the intermediate states accurately. In experimental settings, there is a need to reduce the sampling time for obtaining \emph{reward} signals. One approach was proposed to define a reduced distance metric based on partial state representations at each step, e.g., the distance between the actual partial Bloch vector representation and the target partial Bloch vector representation~\cite{jiang2022robust}. To take it further, one may provide the \emph{reward} signal at the end of the episode and use the \emph{reward} $r_{t<T}=0$ at all intermediate time steps~\cite{mackeprang2019reinforcement,an2021quantum,sivak2022model,sivak2023real}. Considering that a \emph{reward} signal generally involves a measurement process that inevitably disrupts the quantum state, reducing the sampling of \emph{reward} signals mitigates the impact of state collapse (random jumps). While this strategy does offer the advantage of high experimental sample efficiency~\cite{sivak2022model}, it can potentially disrupt the guiding learning process, which traditionally relies on accurate \emph{reward} signals. To address the sparse \emph{reward} signal in partial observation settings, one can introduce an auxillary task to predict \emph{reward} signals~\cite{zhou2023auxiliary}. 

To realize a good tradeoff between accuracy and efficiency when applying RL to quantum systems with partial observation in experimental settings, the capabilities of conventional simulation-based techniques can be leveraged~\cite{sutton2018reinforcement}. 
An effective solution involves a pretraining of the \emph{agent}’s policy within a simulation setting, which equips the \emph{agent} with a better initial point that benefits the subsequent retraining in the real-world setting. 
For example, one can first pre-train the model using the information provided by traditional methods (a \emph{reward} signal suggested from shortcuts to adiabaticity to punish the deviations from linear growth of detuning), 
then fine-tune the models with a different \emph{reward} signal under random systematic errors. This method shows favorable robustness with the increase of the number of $\pi$ flips~\cite{ai2022experimentally}. 
To further improve the robustness of DRL methods in different scenarios, transfer learning can be introduced to bridge the gaps between simulated and real-world \emph{environment}s, ensuring a smoother transition from one to the other.


	
\subsection{RL for quantum error correction}\label{Subsec:QEC}
	
	\begin{figure}
		\centering
		\includegraphics[width=0.48\textwidth]{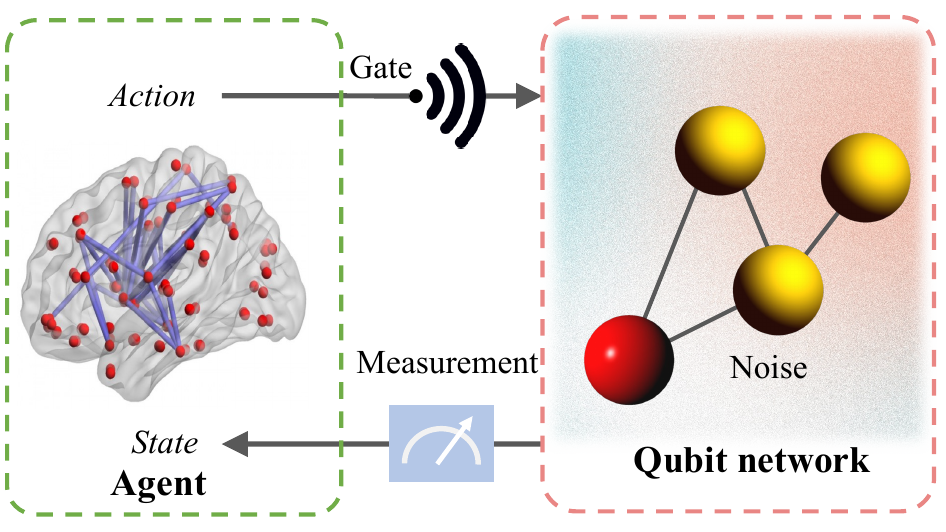}
		\caption{The schematic of protecting quantum memories (qubit network) from the detrimental effects of noise via RL. Given a quantum device consisting of a few qubits, the RL \emph{agent} is required to take \emph{actions} (i.e., making a selection from gate sequences, and execution of measurements). To obtain the optimal effects, the RL \emph{agent} responds to measurement outcomes and collects \emph{reward} signals to guide the RL \emph{agent} towards good  \emph{actions}.}
		\label{fig:DRLforQEC}
	\end{figure}
	
	Within the realm of quantum information and computation, quantum error correction (QEC) stands out as a cornerstone, being widely recognized as the crucial basis for achieving fault-tolerant quantum computation~\cite{terhal2015quantum}. QEC covers a wide range of scenarios, offering a selection of established schemes, such as stabilizer codes~\cite{sivak2022model}. It also has relevance to the broad field of quantum error mitigation techniques. 

	The fundamental objective of QEC is to counteract the intrinsic tendency of a complex system to undergo decoherence. To this end, artificial error-correcting dissipation is established to remove the entropy from the system in an efficient manner by prioritizing the correction of frequent small errors, while not neglecting rare large errors. Let us discuss the particular case of stabilizer-code-based QEC, which involves four steps: encoding, detection, correction, and decoding.  
 

	A complete QEC involves a cooperative process that requires the participation of multiple quantum and classical components. 
	Among the four primary steps, encoding and decoding are two typical procedures that focus on transformations between logical and physical states. The utilization of RL helps guide the \emph{agent} to perform certain \emph{action} in the form of fault-tolerant local deformations of the code, thus benefiting the processes of detection and correction. Fig.~\ref{fig:DRLforQEC} illustrates an example of applying RL to discover an adaptive strategy that protects quantum memory against noise~\cite{fosel2018reinforcement}. The \emph{environment}, in this context, consists of a 
	quantum memory (subject to noise) and its classical
	control system guiding the QEC. In each round of
	interaction, the \emph{agent} receives sensory input from the \emph{environment}, providing information about the current state of the quantum device. The goal of RL is to determine an optimal sequence of \emph{actions} (e.g., quantum gates and measurements) that the \emph{agent} can perform in response to the evolving state of the quantum memory. To achieve this, the \emph{agent} is rewarded if the quantum device has been successfully protected, i.e., if the error rate drops below the specific target. 
	
	Remarkable achievements have been realized in the realm of RL for QEC. One accomplishment is a unified, fully autonomous approach based on policy gradient to discovering QEC strategies from scratch in few-qubit quantum systems subject to arbitrary noise and
	hardware constraints~\cite{fosel2018reinforcement}. The capability of DRL has been well demonstrated by its success in preparing stabilizer states in an oscillator~\cite{sivak2022model}.  {\color{black}A fully stabilized and error-corrected logical qubit has been constructed with quantum coherence substantially longer than that of all the imperfect quantum components involved in the QEC process~\cite{sivak2023real}. The QEC circuit parameters are trained in situ with PPO, ensuring their adaptation to real error channels and control imperfections of the system}. Additionally, a physics-motivated RL variant, known as projective simulation, has been applied to modify a family of surface code quantum memories until a desired logical error rate is reached~\cite{nautrup2019optimizing}. The remarkable generalization of projective simulation has been demonstrated by transferring the experience from a simulated \emph{environment} to different physical setups~\cite{melnikov2017projective}. 
	
	While QEC can be addressed using RL, it is crucial to take into account the specific characteristics, (i) Measurement-based feedback: Apart from typical quantum gates, projective measurements can be included as possible  \emph{actions} that manipulate dynamics of quantum states~\cite{fosel2018reinforcement,sivak2022model}. Those measurement-based  \emph{actions} may introduce discontinuous jumps of quantum states, introducing intricate physical dynamics of quantum systems. (ii) A complete QEC typically involves both recording and correcting errors. It is desirable to design different \emph{reward} signals to distinguish different QEC levels~\cite{fosel2018reinforcement}. For example, one may define a ``protective" \emph{reward} signal to keep the large quantum recovery information among the quantum states~\cite{fosel2018reinforcement}. If one wishes to fully decode the quantum state, a ``recovery" \emph{reward} that considers the overlap between the original state and the recovered state may be designed to encourage operations that can correct errors of quantum states.

	\subsection{Outlook and future directions}\label{Subsec5:summary}

RL provides a promising framework for tackling the challenges of quantum control, enabling the development of effective and adaptive control strategies in quantum systems. Compared to learning-based approaches in Section~\ref{Sec:learningcontrol}, RL methods exhibit the following notable advantages: 1) The introduction of a \emph{reward} signal at each step throughout the entire control pulse, rather than a single ``fitness" value after the control pulse, allows for flexible control of the quantum system (e.g., varied control pulses); 2) Incorporating NNs in RL enables effective optimization of quantum systems under challenging conditions, such as partial observation of quantum systems available; 3) DRL methods aim to learn state-\emph{action} patterns from large-scale data (i.e., the previous experience via trial-and-error), demonstrating improved robustness against errors. 
Despite these advantages, several open questions remain and RL for quantum control deserves further developing.
	
\textbf{High dimensionality}: Exploring quantum systems with high-dimensional state or  \emph{action} spaces typically requires substantial computational resources, e.g., a large number of input or output layers or deep hidden layers among the whole NN design. Scaling up RL algorithms to handle large quantum systems presents significant computational challenges and necessitates the development of efficient algorithms.
	
\textbf{Sample efficiency}: When searching for optimal control in complex quantum systems, performance may be hindered by sparse \emph{rewards}, as early transitions may not achieve a good fidelity thus providing little information to train the RL \emph{agent}. Without additional measures, it may require a large number of training iterations to find an effective control. Utilizing \emph{reward}-shaping techniques that learn \emph{reward} signals from NNs, rather than relying on predefined human-designed \emph{rewards}, may encourage more effective exploration, thus improving sample efficiency. Additionally, incorporating advanced learning strategies such as curriculum learning or transfer learning may enable efficient exploration of quantum systems by reusing knowledge gained from existing experiences. This approach reduces the need to train on a large number of new samples to capture useful state- \emph{action} patterns, thereby enhancing overall efficiency.


\textbf{Real-time implementation}: A key challenge in applying RL to quantum experiments is the disparity between the classical processing timescale and that of quantum systems. To implement control suggested by the RL \emph{agent} on a sub-microsecond timescale, it is crucial to design a low-latency NN architecture allowing for processing data concurrently with its acquisition on hardware, such as field-programmable gate array (FPGA). The reduced latencies in the data processing and analysis in the \emph{agent} are key to real-time control of quantum systems.

\section{Discussion and conclusions}\label{Sec:conclusion}

In this review, we have investigated various quantum tasks, considering the complexity of quantum states and dynamics, as well as the intrinsic probabilistic nature of quantum measurements. Notably, ML techniques are frequently applied to capture information about quantum systems via post-processing routines or to manipulate them toward desired targets through adaptive optimization routines. The advantages of NNs for quantum estimation, learning-based optimization of quantum systems, and RL for quantum systems highlight the significant power of ML in addressing quantum challenges. With the recent progress in ML, it is highly desirable to leverage these methods to tackle complex quantum problems effectively. Moreover, the combination of ML and quantum technologies leads to the emerging area of quantum machine learning, opening new application opportunities in quantum estimation and control, and driving advancements in the fields.

	\bibliographystyle{ieeetr}
	\bibliography{bib/mybib}
	
\end{document}